\documentclass[preprint]{aastex63}
\usepackage{mathtools}

\shorttitle{Low-Redshift FeLoBALQs III}
\shortauthors{Choi et al.}

\usepackage{CJKutf8}
\begin{document}
\begin{CJK}{UTF8}{}
\CJKfamily{mj}

\title{The Properties of Low-Redshift FeLoBAL Quasars: III. The
  Location and Geometry of the Outflows}
  
\correspondingauthor{Hyunseop Choi}
\email{hyunseop.choi@ou.edu}

\author[0000-0002-3173-1098]{Hyunseop Choi (최현섭)}
\affiliation{Homer L.\ Dodge Department of Physics and Astronomy, The
  University of Oklahoma, 440 W.\ Brooks St., Norman, OK 73019, USA}

\author[0000-0002-3809-0051]{Karen M.\ Leighly}
\affiliation{Homer L.\ Dodge Department of Physics and Astronomy, The
  University of Oklahoma, 440 W.\ Brooks St., Norman, OK 73019, USA}

\author{Collin Dabbieri}
\affiliation{Department of Physics and Astronomy, Vanderbilt
  University, 2301 Vanderbilt Place, Nashville, TN 37235-1807}
\affiliation{Homer L.\ Dodge Department of Physics and Astronomy, The
  University of Oklahoma, 440 W.\ Brooks St., Norman, OK 73019, USA}

\author{Donald M.\ Terndrup}
\affiliation{Department of Astronomy, The Ohio State University, 140
  W. 18th Ave., Columbus, OH 43210}
\affiliation{Homer L.\ Dodge Department of Physics and Astronomy,
  The
  University of Oklahoma, 440 W.\ Brooks St., Norman, OK 73019, USA}

\author{Sarah C.\ Gallagher}
\affiliation{Department of Physics \& Astronomy, The University of
  Western Ontario, London, ON, N6A 3K7, Canada}
\affiliation{Canadian Space Agency, 6767 Route de l'A\'eroport,
  Saint-Hubert, Quebec, J3Y BY9}
\affiliation{Institute for Earth and Space Exploration, The
  University of Western Ontario, London, ON, N6A 3K7, Canada}
\affiliation{The Rotman Institute of Philosophy, The University of
  Western Ontario, London, ON, N6A 3K7, Canada}

\author{Gordon T.\ Richards}
\affiliation{Department of Physics, Drexel University, 32 S. 32nd St.,
  Philadelphia, PA 19104}



\begin{abstract}
We present continued analysis of a sample of low-redshift iron low-ionization broad absorption-line
quasars (FeLoBALQs).  \citet{choi22} presented {\it SimBAL} spectral
analysis of BAL outflows in 50 objects.  \citet{leighly22}
analyzed optical emission lines of 30 of those 50
  objects and found that they are characterized by either a high
accretion rate ($L_\mathrm{Bol}/L_\mathrm{Edd}>0.3$) or low accretion
rate ($0.03<L_\mathrm{Bol}/L_\mathrm{Edd}<0.3$). We report
that the outflow velocity is inversely correlated 
with the BAL location among the high accretion rate objects, with the 
highest velocities observed in the parsec-scale outflows. In
contrast, the low Eddington ratio objects showed the opposite trend. We confirmed the
known relationship between outflow velocity and
$L_\mathrm{Bol}/L_\mathrm{Edd}$, and found that the scatter plausibly
originates in the force multiplier (launch radius) in the low (high)
accretion rate objects.  A log volume filling factor between $-6$ 
and $-4$ was found in most outflows, but was as high as $-1$
for low-velocity compact outflows. We investigated the relationship
between the observed  [\ion{O}{3}] emission and that predicted from
the BAL gas. We found that these could be
reconciled  if the emission-line covering fraction depends on
Seyfert type and BAL location.
The difference between the predicted and observed [\ion{O}{3}] luminosity is
correlated with the outflow velocity, suggesting that [\ion{O}{3}]
 emission in high Eddington ratio objects may be broad and hidden
 under \ion{Fe}{2} emission.   
We suggest that the physical differences in the
outflow properties as a function of location in the quasar and
accretion rate point to different formation, acceleration, and
confinement mechanisms for the two FeLoBALQ types.

\end{abstract}



\section{Introduction} \label{intro}

The broad and blueshifted \ion{C}{4}$\lambda\lambda 1548,1551$ lines
observed in 
broad absorption-line quasars (BALQs) reveal the unambiguous signature
of outflows.  Therefore, BALQs may be important sources of quasar
feedback in galaxy evolution.  In this context, an important parameter
is the ratio of the kinetic luminosity ($L_\mathrm{KE}$) to bolometric
luminosity ($L_\mathrm{Bol}$), because it has been shown that if
$L_\mathrm{KE}/L_\mathrm{Bol}$ exceeds 0.5--5\%
\citep{dimatteo05,he10} then sufficient energy is available to
regulate star formation and therefore produce the distribution of
galaxies that we see today.  

Determining $L_\mathrm{KE}/L_\mathrm{Bol}$ in the general population
of BALQs is challenging.  The measurement of $L_\mathrm{KE}$ requires
constraint of key physical parameters of the outflowing gas including
the column density $N_H$, the velocity, and the location of the
outflow, as well as an estimate of $\Omega$, the fraction of the full $4\pi$
sterradians that is covered by the outflow.  The velocity can be
estimated from the blueshift of the absorption lines, and $\Omega$ is
estimated from the incidence of BAL quasars in the population.  The
column density and location require measurement of the
photoionization properties of the outflowing gas.  These properties
can be inferred directly from measurements  of the optical depths of
absorption lines in the spectrum, but only when the lines are
relatively narrow and line blending is not severe.  This approach has
been used on some tens of spectra \citep[e.g.,][and references
  therein]{arav20,miller20}. 

However, this  type of analysis cannot be done on spectra in which
line blending is significant, i.e., most of the $\sim$30,000 BAL quasar
spectra present in the SDSS archive \citep[DR14Q;][]{paris18}.
It is easy  to understand why we cannot generalize results from 10s of
spectra with narrow absorption lines to the quasar
population in general.  The kinetic luminosity depends 
principally on the outflow velocity, since that factor enters the
equation for $L_\mathrm{KE}$ to the cubed power.  Moreover,
observations show that BAL velocities have a tremendous range, from
nearly zero velocity or inflow in a few cases to more than 30,000$\rm
\, km\, s^{-1}$.  At high velocities, blending is usually significant.
Samples of BAL quasars repeatedly show a relationship 
between the BAL outflow velocity and the luminosity of the quasar
\citep[e.g.,][]{laor02, ganguly07}, but the dependence is complicated,
with the relationship showing an upper limit envelope rather than a
one-to-one dependence.  So, we can expect that a powerful quasar
located at $z=2$--3 that may have a bolometric luminosity exceeding
$10^{48}\rm \, erg s^{-1}$ will have much different outflow properties
and $L_\mathrm{KE}/L_\mathrm{Bol}$ than a Seyfert luminosity object at
$z=0.5$--1.5.   

It is also very plausible that $L_\mathrm{KE}/L_\mathrm{Bol}$
depends on the Eddington ratio $L_\mathrm{Bol}/L_\mathrm{Edd}$
of the quasar.  The terminal velocity of an accelerated outflow
depends on the magnitude of the acceleration but also on the
deceleration due to gravity provided by the supermassive black hole.
Also, we expect the geometry and physical conditions of the central
engine to change with this parameter \citep[e.g.,][]{gp19}.  Finally,
we infer profound dependence of the broad emission lines on
$L_\mathrm{Bol}/L_\mathrm{Edd}$ and ask why the absorption lines
  should  not also depend on this factor.
The bottom line  is that
it is probably not reasonable to expect that a single 
value of $L_\mathrm{KE}/L_\mathrm{Bol}$ should be applicable for all UV
outflows, but rather that this parameter should depend on (at least)
the luminosity (equivalently the black hole mass) and the Eddington
ratio.  

A first step in making progress on the relationships between
  fundamental quasar parameters and outflow properties would be to
analyze broad absorption-lines from many more objects representing the
full range of the BAL phenomena.  This step requires a method to
handle line blending.
Here, we use the term ``line
blending'' to mean that the velocity width of the broad absorption-lines is sufficiently large that different absorption lines blend
together regardless of the spectral resolution.
That problem is now approachable using the
novel spectral synthesis code {\it SimBAL}.
The {\it SimBAL} methodology is described in \citet{leighly18}.  {\it SimBAL} uses a forward modeling method.
It creates  synthetic spectra parameterized by photoionization conditions of the outflowing gas, and then compares them with the observed spectrum using a Markov Chain Monte Carlo
method \citep[{\tt emcee};][]{emcee}.  Additional examples of the use
of {\it SimBAL} can be found in \citet{leighly19} and \citet{choi20}.
The {\it SimBAL} analysis of the objects described in this paper is
found in \citet{choi22}.

The second thing we can do is to try to understand the physics of BAL
outflows.  If that could be done, then we could predict the outflow
properties in a given quasar, given its (say) luminosity and Eddington
ratio.
Despite fifty years of study, it is still not known how
these outflows are accelerated, what confines the BAL ``clouds'', or
what the origin of the absorbing gas is.  In short, the same questions
that were posed in the '80s and '90s are still unanswered and are
still relevant today.   

Some of these questions can be addressed by studying the rest-frame
optical emission lines in BAL quasars.  First, the $H\beta$$\lambda 4863$\AA\/ emission
line region yields an estimate of the black hole mass.   Black hole
mass estimates are also available using the \ion{Mg}{2}$\lambda\lambda
2796,2804$ 
\citep[e.g.,][]{bahk19} and \ion{C}{4} \citep[e.g.,][]{coatman17}, but
those lines may be significantly absorbed in BAL quasars, a fact that 
adds significant uncertainty.
In contrast, it is rare to observe absorption in Balmer lines and even when they are present in the spectra, the Balmer emission lines (e.g., H$\beta$) generally are not significantly absorbed and can be easily studied\citep[e.g.,][]{schulze18}.
The H$\beta$ region also includes
[\ion{O}{3}]$\lambda 4960, 5008$ and many thousands of lines in the
\ion{Fe}{2} pseudocontinuum emission, and together with H$\beta$,
these parameters are thought to reflect the physical conditions of the
central engine through a pattern of behavior known as Eigenvector 1
\citep{bg92} that is widely considered to be a probe of the
Eddington ratio.
Eigenvector 1 is repeatedly found in principal
  components analysis (PCA) of optical rest-frame quasar spectra
  \citep[e.g.,][]{grupe04,ludwig09,wolf20}.
Thus, we might expect that, just as we observed the
BAL outflow velocity to depend on the quasar luminosity, we should
observed BAL properties to depend on the Eigenvector 1 properties.  

BAL quasars are divided into three types based on the absorption lines
present in the spectrum.
\ion{C}{4} is the most commonly observed line, observed in 10-26\% of optically
selected quasars \citep{tolea02, hf03, reichard03, trump06, knigge08,  gibson09}. 
  Objects that only have this line
plus other high-ionization lines such as \ion{N}{5}$\lambda\lambda
1239,1242$ and \ion{Si}{4}$\lambda\lambda 1394, 1403$ 
are called high-ionization broad absorption-line quasars (HiBALQs).  About 1.3\% of quasars have broad
\ion{Mg}{2} absorption \citep{trump06}; these are called
low-ionization broad absorption-line quasars (LoBALQs). About 0.3\% of
quasars also have absorption from \ion{Fe}{2}, and these are called
iron low-ionization broad absorption-line quasars
\citep[FeLoBALQs,][]{trump06}.
These objects are observed much less frequently than the other two
categories, but to some extent, their rarity can be attributed to the 
difficulties in detection these objects, since they may lack strong
emission lines due to absorption.  Furthermore, their spectral energy
distributions (SEDs) show the reddest  colors among BAL quasars which
suggests that they may represent a short-lived stage in quasar
evolution where the quasar expels its cocoon of gas and dust
(“blowout” phase; e.g., Sanders et al. 1988), transitioning from the
ultraluminous infrared galaxy (ULIRG) phase \citep[e.g.,][]{urrutia09,
  glikman17, glikman18}.    Finally, the physical conditions of the
outflow gas can be particularly well constrained using the thousands
of \ion{Fe}{2} absorption lines.

This paper is the third in a sequence of four papers reporting the
results of a comprehensive analysis of a sample of FeLoBAL
quasars. Paper I \citep{choi22} describes the sample of 50
low redshift ($0.66 \leq z \leq 1.63$) FeLoBALQs and the {\it
  SimBAL} spectral synthesis 
analysis of their  absorption lines.  That work represents an increase
by a factor of five of FeLoBALQs with detailed photoionization
analysis. We measured the  velocity and velocity width, the ionization
parameter, density, column density and covering fraction  directly
from the spectra.  We extracted the location of the outflow, as well
as the mass outflow rate and energetics of the outflow. We found that
FeLoBAL outflows cover a large range of  outflow locations in a
quasar, spanning $\log R $ between 0 and 4.4 [pc].  While many of the
troughs were well described  by a single outflow component
characterized by a single ionization parameter and density, about 20\%
of the objects showed evidence for multiple troughs, where the higher
velocity components generally had higher ionization parameters.  Among
these objects, several special classes of BAL outflows were found.
Overlapping trough objects \citep[e.g.,][]{hall02} show broad
absorption troughs that blanket the near-UV shortward of \ion{Mg}{2}.
All of these objects were found to have compact outflows with $\log R
< 1$ [pc].  Among these objects, we discovered a new type of FeLoBALQ.
Dubbed ``loitering'' outflow FeLoBALQ, these objects generally have
narrow lines and low velocities, and are also compact  with $\log
R < 1$ [pc].  They are distinguished from the other  overlapping trough
objects not only by their velocity width but also their tendency to
have higher ionization parameter and higher density gas
\citep[Fig.\ 6][]{choi22}, which leads
to opacity from many high excitation states.  The  outflowing gas in
about half of these objects occults only the continuum emission
region, but not the broad emission lines.  The remaining objects were
located farther from the quasar and generally did not have extreme
properties.   We also found that there was a significant
correlation between the color of the quasar UV--NIR SED and the
outflow strength where the quasars with redder SEDs have more
energetic outflows.
Finally, we discussed how the potential
acceleration mechanisms and the origins of the FeLoBAL winds may
differ for outflows at different locations in the quasars.

Thirty of the fifty FeLoBALQs analyzed in \citet{choi22} have
sufficiently low redshift that the H$\beta$ emission-line region is
present at the red end of the SDSS spectra. Paper II \citet{leighly22}
describes the rest-frame optical emission-line analysis 
of this subsample, along with a 132-object 
comparison sample of non-BAL quasars.  The principal result of that
paper is that the FeLoBALQs are divided into two classes based on
their emission-line properties, and their emission line properties are
distributed differently than those of the comparison sample.
Specifically, FeLoBALQs were characterized by either weak
  \ion{Fe}{2} relative to H$\beta$ and strong [\ion{O}{3}], or strong
  \ion{Fe}{2} relative to H$\beta$ and weak [\ion{O}{3}], and
  intermediate values were avoided.
Further analysis  revealed that the  emission line properties reflect the
accretion rate relative to the Eddington limit.  Therefore, 
FeLoBAL quasars at low redshift are characterized by either high
Eddington ratios (typically greater than 1), a result that agrees with previous analysis of BAL
quasars \citep{yw03,   boroson02, runnoe13}, or low Eddington ratios (typically less than 0.1),
a new result, but are uncommon at the intermediate Eddington ratios that
are the most prevalent in the comparison sample.    The fact that their
emission-line properties are different than unabsorbed quasars shows
that among low-redshift and low-luminosity objects, FeLoBAL and
non-BAL quasars do not have the same parent sample.  This result is
different than has been recently  reported for high-ionization broad
absorption line (HiBAL) quasars which have higher luminosities
\citep{rankin20}.   

This paper combines the {\it  SimBAL} results from  \citet{choi22} and
emission-line analysis results of the 30-object subsample from
\citet{leighly22}  to search for relationships between the
properties of the outflowing gas and the properties of the central
engine.  \citet{choi22} demonstrated that the BAL outflow
velocity is related to the bolometric luminosity in the 50-object
sample, as has been found previously
\citep[e.g.,][]{laor02,ganguly07,fiore17}.  However, {\it SimBAL}
delivers 
quite a bit more information characterizing the outflow than typical
BAL analyses, including the parameters describing the ionization
state, density, column density, and covering fraction, but also the
location of the outflow.  We therefore take the first steps in
tackling the question posed above and examine how the location,
geometry, confinement and other properties of the outflowing gas
depend on the global quasar properties such as the luminosity and
Eddington ratio. 

The final paper in the series, Paper IV (Leighly et al.\ in prep.),
includes an analysis of the the broad-band optical/IR properties and
discusses the potential implications for quasar evolution scenarios.   

This paper is organized as follows.  In \S\ref{data} we briefly
describe the data extracted from \citet{choi22} and \citet{leighly22}.
We principally focused on the 30-object $z<1$ subsample 
for which the H$\beta$ region is available and the analysis described
in \citet{leighly22}, although we also explored the volume
filling factor in the full sample.  In \S\ref{dist_comp}, we used the
$E1$ parameter defined in \citet{leighly22} to divide the
FeLoBAL quasars into high and low accretion rate objects and then
compared the {\it SimBAL} properties of the two classes.  We also
correlated the {\it SimBAL} parameters with one another, and with the
optical emission-line and global parameters.  Finally, we investigated
the relationship between the properties of the observed [\ion{O}{3}]
emission lines and the [\ion{O}{3}] emission predicted to originate in
the BAL gas.  \S\ref{discussion} presents a summary of the results.

We used cosmological parameters $\Omega_\Lambda$=0.7,
$\Omega_M=0.3$, and $H_0=70\rm\, km\, s^{-1}\, Mpc^{-1}$, unless
otherwise specified.

\section{Data} \label{data}

The data used in this paper are described in detail in Papers I
\citep{choi22}  and
II \citep{leighly22}l. The {\it SimBAL} model-fitting results
were drawn from Paper I, and the optical emission-line modeling and
global properties were taken from Paper II.  Those parameters are
described briefly in the next sections.  

\subsection{{\it SimBAL} Parameters} \label{simbal}

The {\it SimBAL} model fits for the sample are given in 
\citet{choi22}, and the details can be found in that paper
(Tables 2 and 3).  We extracted the following parameters from those
results: the ionization parameter $\log U$, the gas density $\log n \, \rm
[cm^{-2}]$, the broad absorption line velocity $V_\mathrm{off}\rm \, 
(km\, s^{-1})$,  the broad absorption line velocity width
$V_{width}\rm \,  (km\, s^{-1})$\footnote{The representative offset
  velocity is the   median value of the MCMC chain after weighting by
  the true opacity. The true opacity is distinguished from
    the apparent opacity in that it takes into account partial
    covering of continuum and emission-line emitting region.  The
    apparent opacity is extracted directly from the spectrum.; see \citet{choi22} \S~4.1.3 for the
  definition.  The minimum and   maximum velocities and the velocity
  width were estimated from the 
  90\% transmittance level from the model \ion{Mg}{2}$\lambda 2796$
  velocity profile.}, the total column density integrated over
  the BAL feature $\log N_{H}\, \rm [cm^{-2}]$,
the covering fraction parameter $\log a$, and the radius of the
outflow $\log R$ [pc].

Several additional and derived parameters were also produced that are
not reported in tables in  \citet{choi22}.  These include
the largest and smallest outflow velocities $V_{max}$, $V_{min}$,
force multiplier (the ratio of the total opacity to the electron
scattering opacity; FM), the thickness of the outflow $\Delta R$ [pc],
the filling factor $\log \Delta R/R$, the net outflow rate $\log
\dot{M}\rm \, [M_{\odot}\,   yr^{-1}]$ (per component), the net
kinetic luminosity $\log L_\mathrm{KE}\, \rm [erg\, s^{-1}]$, and the
ratio of the net kinetic luminosity to the bolometric luminosity.  The
thickness of the outflow  $\Delta R$ is the ratio 
of the total column density and the density.  Making the simple
assumption that the azimuthal size of the outflowing gas is comparable
to the thickness of the gas (i.e., the gas is distributed into
  individual clouds and the clouds are approximately spherical), and
using the radius of the accretion 
disk at 2800\AA\/ described in \citet{leighly22},  we
computed the log of the number of clouds required to cover the
continuum emission region (see \S\ref{covfrac} for more details). 

Finally, we used the parameters from the {\it SimBAL} best fitting
solution to compute the luminosity of the predicted [\ion{O}{3}]
emission line assuming an emission-line global covering fraction of $0.1$
(discussed in \S~\ref{oiiiemission}). We also computed a parameter
called the covering fraction correction, described in that section,
which parameterizes the comparison of the predicted  [\ion{O}{3}]
luminosity from the wind with the observed [\ion{O}{3}] luminosity.   

\subsection{Optical and Global Parameters} \label{opt_glob}

The following optical emission-line parameters were taken from Table 1 
of \citet{leighly22}.  We used the H$\beta$ FWHM and
equivalent width to parameterize the H$\beta$ line. \citet{leighly22}
also defined a  
parameter called the H$\beta$ deviation that measures the
systematically broader H$\beta$ FWHM observed among the FeLoBAL
quasars compared with unabsorbed quasars. \ion{Fe}{2} was 
parameterized using  R$_\mathrm{FeII}$ defined as the ratio of the
\ion{Fe}{2} equivalent width in the range 4434--4684 \AA\/ to the broad
H$\beta$ equivalent width \citep[e.g.,][]{shen_ho_14}.  The
[\ion{O}{3}] emission line was  parameterized using the equivalent
width and luminosity, along with profile parameters $v_{50}$ and
$w_{80}$ defined according to the  prescription of
\citet{zg14}. Briefly, from the normalized cumulative function of the
broad [\ion{O}{3}] model profile, the velocities at 0.1, 0.5, and 0.9
were identified. The velocity at 0.5 is assigned to $v_{50}$, and
$w_{80}$ is the  difference between the velocities at 0.1 and 0.9.
\citet{leighly22} also fit the FeLoBAL objects and the  unabsorbed
objects with eigenvectors created from the continuum-subtracted
spectra of the unabsorbed objects \citep[\S2.3,][]{leighly22}.
The fit coefficients for the first 
four  eigenvectors (SPCA1--4) serve  as a  parameterization of the
spectra. The first eigenvector displays the relationship between
  the strength of \ion{Fe}{2} and [\ion{O}{3}] that is commonly found
  \citep[e.g.,][]{grupe04, ludwig09, wolf20}, and none of the other
  ones display any particular anomalies. We estimated the
bolometric luminosity using the rest-frame 
flux density at 3 microns and a bolometric correction of 8.59 
\citep{gallagher07}.
BAL quasars tend to be reddened
  \citep[e.g.,][]{krawczyk15}, and evidence for reddening is present
  in this sample \citep[$0\lesssim E(\bv)\lesssim0.5$, Figure~20 in][]{choi22,leighly_prep_paper4}.
Therefore, we used the   3-micron luminosity
  density as   representative, rather than the luminosity in   the
  optical or UV.
The black hole mass and Eddington ratio were
computed using standard methods and as described in \citet[][\S
  2.1]{leighly22}, and the calculation of the location of the
2800\AA\/ emission from the accretion disk follows the method used in 
\citet{leighly19}. 

Finally, we continue to use the $E1$ parameter, which was defined in
\S 3.1 of \citet{leighly22}.   This parameter is a function
of the measured values of R$_\mathrm{FeII}$ and the [\ion{O}{3}]
equivalent width.
As described in \S 3.1 of \citet{leighly22},
  we normalized and scaled R$_\mathrm{FeII}$ and the 
  [\ion{O}{3}] equivalent width of the 132-object comparison sample,
  and then derived the bisector line. 
  We   performed a coordinate rotation so that $E1$ is a parameter
  that lies along the   bisector.  Our $E1$ parameter is therefore
  related to the \citet{bg92} Eigenvector 1.  \citet{bg92} performed a
  principal components analysis of emission line properties in the
  vicinity of H$\beta$ and found that the variance is dominated by an 
anticorrelation between R$_\mathrm{FeII}$ and the
  [\ion{O}{3}] equivalent width.  Moreover, Eigenvector 1 has been
  shown to dominate the variance in quasar emission-line properties
\citep[e.g.,][]{francis92,bg92,brotherton94,corbin96,wills99,sulentic00, 
  grupe04,yip04,wang06,ludwig09,shen16}.
$E1$ is very strongly correlated with SPCA1
($p=6.5\times 10^{-15}$ and $p=10^{-50}$ for the FeLoBALQs and
unabsorbed comparison sample that was analyzed in parallel,
respectively).  It is also correlated 
with the Eddington ratio ($p=1.2\times 10^{-5}$ and $p=9.3\times
10^{-7}$ for the FeLoBALQs and unabsorbed comparison sample,
respectively).
Here, $p$ is a measure of the statistical
  significance  of the correlation.  More specifically, it is a
  probability that the observed correlation could have been produced
  from draws from two uncorrelated samples \citep[e.g.,][]{bevington}.
A low (negative) value of the $E1$ parameter corresponds to a low
accretion rate, while a high (positive) value of the $E1$ parameter
corresponds to a high accretion rate. The 90\% range of $\log
L_\mathrm{bol}/L_\mathrm{Edd}$ among the sample is -1.4 to 0.84
\citep[Fig.\ 9][]{leighly22}.  As discussed in \S 3.3 of
\citet{leighly22}, the apparent bimodal 
distribution of FeLoBALQs in the $E1$ shows that there are two
types of FeLoBALQ: objects with $E1<0$ are characterized by low
Eddington ratio, and objects with $E1>0$ are characterized by high
Eddington ratio.  We divided the FeLoBALQs into two groups based on
this parameter, where the dividing line $E1=0$ corresponds to $\log
L_\mathrm{Bol}/L_\mathrm{Edd}=-0.5$, i.e., an Eddington ratio of about
0.3.  Throughout this paper, we use a consistent coloring scheme to
denote $E1$: red (blue) corresponds to $E1<0$ and low Eddington ratio
($E1>0$ and high Eddington ratio).

\section{Distributions and  Correlations Between Optical and BAL Outflow Parameters}\label{dist_comp}

\subsection{Distributions}\label{distributions}

\citet{leighly22} presented comparisons of the distributions
of the emission-line and derived properties of the FeLoBAL quasars 
with those of the comparison sample of unabsorbed quasars using
cumulative distribution plots.   We applied the  two-sample
Kolmogorov-Smirnov (KS) test, and the two-sample Anderson-Darling (AD)
test.
The KS test reliably tests the difference
between two distributions when the difference is large at the median
values, while the AD test is more reliable if the differences lie
toward the maximum or minimum values (i.e., the median can be
the same, and the distributions different at larger and smaller
values)\footnote{E.g.,
  https://asaip.psu.edu/articles/beware-the-kolmogorov-smirnov-test/}.
We also compared the samples divided by the sign of the $E1$
parameter. Note that the values of the $E1$ parameter and plots of the
spectral model fits are given in \citet{leighly22}.    In this paper,
we present the comparisons of the {\it SimBAL} parameters for the
FeLoBAL quasars segregated by $E1$ parameter (Table~\ref{tab_distributions}).

\startlongtable
\begin{deluxetable}{lCC}
\tablecaption{{\it SimBAL} $E1<0$ versus $E1>0$ Parameter Comparisons\label{tab_distributions}}
\tablehead{
\colhead{Parameter Name} & \colhead{KS\tablenotemark{a}} &
\colhead{AD\tablenotemark{b}} \\
& Statistic / Probability & Statistic / Probability \\
}
\startdata
$\log U$ &  0.21 / 0.28 & 0.64 / 0.18 \\
$\log n$ [cm$^{-3}$] & 0.24 / 0.57 & -0.07 / >0.25 \\
V$_\mathrm{offset}$ (km s$^{-1}$) & 0.50 / 0.016 & 4.4 / 5.9\times10^{-3} \\
V$_\mathrm{max}$ (km s$^{-1}$) & 0.50 / 0.014 & 4.2 / 7.0\times 10^{-3} \\
V$_\mathrm{min}$ (km s$^{-1}$) & 0.55 / 4.7\times
  10^{-3} & 5.0 / 3.6\times 10^{-3} \\
V$_\mathrm{width}$ (km s$^{-1}$) &  0.30 / 0.33 & 0.67 /
0.17 \\
$\log a$ &  0.36 / 0.14 & 1.1 / 0.11 \\
Net $\log$ N$_\mathrm{H}$ [cm$^{-2}$] &  0.33 / 0.21 &
0.094 / > 0.25 \\
$\log$ Force Multiplier  & 0.35 / 0.17 & 0.32 / >0.25 \\
$\log$ R [pc] & 0.26 / 0.45 & -0.10 / >0.25 \\
$\Delta$ R [pc] & 0.46 / 0.032 & 3.4 / 0.013 \\
$\log$ Volume Filling Factor  & 0.51 / 9.8\times 10^{-3} &
5.7 / 2.0\times 10^{-3} \\
$\log$ Number of Clouds & 0.46 / 0.032 & 4.7 / 4.6\times 10^{-3} \\
Net \.M (M$_\odot$ yr$^{-1}$) & 0.31 / 0.28 &
0.084 / > 0.25 \\
L$_\mathrm{KE}$ [erg s$^{-1}$] & 0.42 / 0.10 & 1.4 / 0.09 \\
L$_\mathrm{KE}$/L$_\mathrm{Bol}$ & 0.30 / 0.41 & 0.37 / 0.24 \\
Predicted [OIII] Luminosity & 0.26 / 0.49 & 0.56 / 0.19 \\
Covering Fraction Correction & 0.58 / 2.1\times 10^{-3} & 9.3 / < 0.001 \\
\enddata
\tablenotetext{a}{The Kolmogorov-Smirnov Two-sample test.  Each entry
  has two numbers:  the first is the value of the statistic, and the
  second is the probability $p$ that the two samples arise from the same
  parent sample.  Bold type indicates entries that yield $p<0.05$.}
\tablenotetext{b}{The Anderson-Darling Two-sample test.  Each entry
  has two numbers:  the first is the value of the statistic, and the
  second is the probability $p$  that the two samples arise from the same
  parent sample.  Note that the implementation
  used does not compute a probability   larger than 0.25 or smaller than
  0.001. Bold type indicates entries that yield $p<0.05$.}
\end{deluxetable}

\begin{figure*}[!t]
\epsscale{1.0}
\begin{center}
\includegraphics[width=5.5truein]{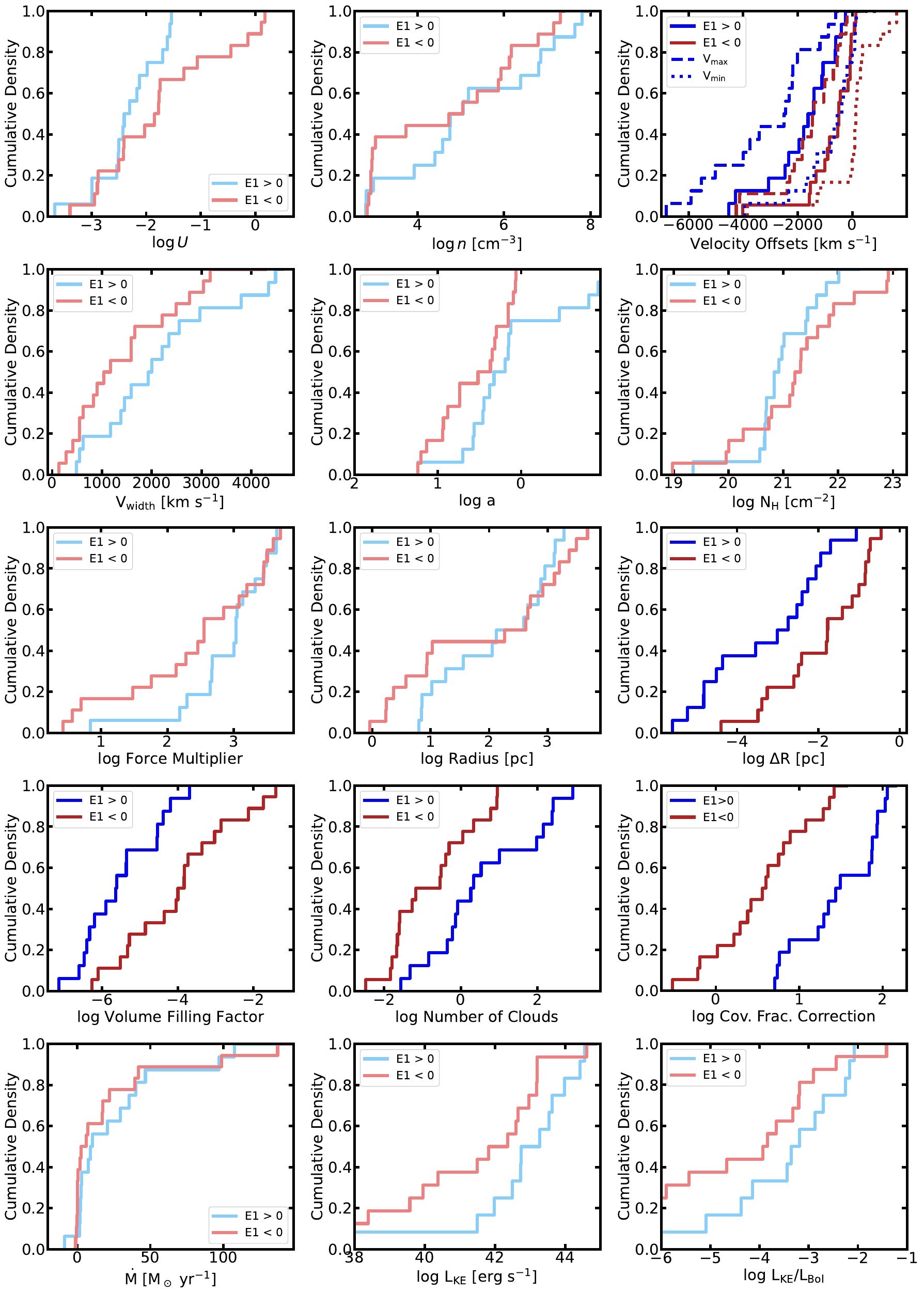}
\caption{The cumulative distributions of seventeen of the {\it SimBAL}
  parameters segregated by their $E1$ parameter value. Distributions
  that are significantly  different ($p<0.05$;
  Table~\ref{tab_distributions}) are shown in dark red   and  dark blue,
  while    distributions that are not significantly different are
  shown in a pale hue of the same color.    All of the
  outflow velocity parameters ($V_\mathrm{off}$, $V_\mathrm{max}$,
  $V_\mathrm{min}$) are shown in the top-right panel.    Of the {\it
    SimBAL}-related parameters, those three, the   thickness of the
  absorber, the log filling factor, the number of   clouds covering
  the continuum   emission-line region, and the   covering fraction
  correction (discussed in \S~\ref{oiiiemission}) 
  show statistically different distributions between the $E1>0$ and
  $E1<0$ subsets.    \label{simbal_dist}} 
\end{center}
\end{figure*}

The results for the {\it SimBAL} parameters are shown in
Fig.~\ref{simbal_dist}, and the statistics are given in
Table~\ref{tab_distributions}.    The
parameters that exhibit statistically significant differences between
the  $E1>0$ and $E1<0$ groups are all three  velocity outflow 
parameters ($V_\mathrm{off}$, $V_\mathrm{max}$,  and
$V_\mathrm{min}$), the thickness of the outflow $\Delta R$, the log
filling factor, the log of the number of clouds covering the continuum
emitting source, and the covering fraction correction factor.  The
$E1>0$   objects show systematically larger opacity-weighted outflow
 velocities than the $E1<0$ objects, with median values $-1520\rm \,
 km\, s^{-1}$ and $-500\rm \, km\, s^{-1}$, respectively. For the
 $E1<0$ objects, the outflows are thicker, have larger filling
 fractions, and fewer\footnote{A median number of
   clouds less than zero might be interpreted as an approximately
   continuous outflow.  Alternatively, a single cloud, if in the line
   of sight, would completely cover the continuum source.} are
 required cover the continuum source (median 
 $\log  \Delta R= -1.8$ [pc], $\log$ volume filling fraction $=-3.8$, and
 $\log$ number of clouds $=-0.54$).  The
 $E1>0$ objects have  
thinner outflows,  smaller volume filling fractions, and more clouds
are necessary to cover the continuum emission ($\Delta R= -2.4$ [pc],
$\log$ volume filling fraction $=-5.6$, and  $\log$  number of
clouds $=0.34$).  These properties are discussed further in
\S~\ref{covfrac}.  The covering fraction correction factor is lower
for the $E1<0$ than for the $E1>0$ objects (median values of 0.6 and
1.5, respectively); this parameter was defined in \S\ref{simbal} and
is  discussed in detail in \S~\ref{oiiiemission}.  

The ionization parameter, $\log a$, and the force multiplier  show
suggestions of differences.  Specifically, while the highest
ionization parameter represented by $E1>0$ objects is $\log U=-0.5$,
25\% of $E1<0$ objects have ionization parameters larger than that
value (Fig.~\ref{simbal_dist}).  Likewise, lower values of force
multiplier are dominated by $E1<0$ objects (Fig.~\ref{simbal_dist}). In
addition, the kinetic luminosity $L_\mathrm{KE}$ and the ratio of the
kinetic to bolometric luminosity are consistently lower for the $E1<0$
objects (Fig.~\ref{simbal_dist}).

\subsection{Correlations}\label{correlations}

In \citet{leighly22}, we examined the relationships among
the optical and derived parameters such as the bolometric luminosity.  In this
paper, we compare those parameters with the {\it SimBAL} parameters,
both for the full sample, and for the sample segregated by $E1$
parameter.  We used the Spearman-rank correlation for our comparisons. 

\begin{figure*}[!t]
\epsscale{1.0}
\begin{center}
\includegraphics[width=3.00truein]{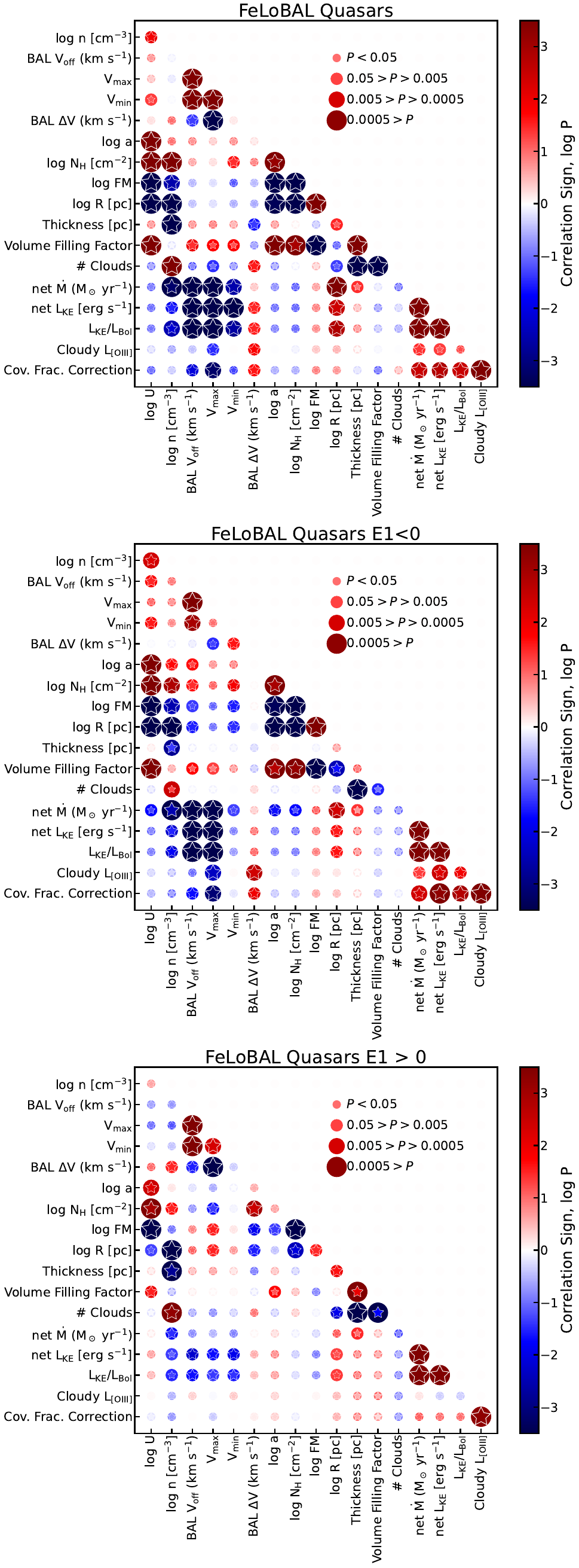}
\caption{The results of the Spearman rank correlation analysis for the
  eighteen {\it SimBAL} parameters. The stars show the results for a
  Monte Carlo
  scheme to estimate the effects of the errors; see \citet{leighly22}
  for details. The top plot shows the results for the
  whole sample, while the middle and bottom plots show the results
  divided by $E1$ parameter.  Of particular interest is the weak
  but significant anticorrelation between the velocity parameters and
  location of the outflow for $E1<0$ and the correlation for the same
  parameters for $E1>0$.     \label{correlation_simbal}}  
\end{center}
\end{figure*}

We first correlated the {\it SimBAL} parameters with one another.  A
correlation analysis of the {\it SimBAL} results for the full
50-object sample is given \S~6.1 of \citet{choi22}.   Here
we considered only the low-redshift subsample.  The results are shown
in  Fig.~\ref{correlation_simbal}.    The plots represent the log of
the $p$ value for the correlation, where the sign of the value gives the
sense of the correlation.  That is, a large negative value implies a
highly significant anticorrelation.    

Parameter uncertainties were propagated through the correlations
using a Monte Carlo scheme.  We made 10,000 normally distributed draws
of each parameter, where the distribution was stretched to the size of
the error bar.  Asymmetrical errors were accounted for by using a
split-normal distribution (i.e., stretching  the positive draws
according to the positive error, and the negative draws according to
the negative error).  We chose $p<0.05$ as our threshold for
significance. In most cases, taking the errors into account did not
dramatically change  the significance of a correlation, if present.  

There are several extremely strong correlations among the {\it
  SimBAL} parameters (Fig.~\ref{correlation_simbal}) that were also
observed for the full sample \citet{choi22}.  Since we
discuss only FeLoBALQs in this paper, most gas columns extend beyond
the hydrogen ionization front\footnote{The hydrogen ionization front is
  the location in a slab of photoionized gas where the
  hydrogen-ionizing photon flux is exhausted.  In the context of
  \ion{H}{2} regions, it is the location of the Str\"omgren sphere.} in 
order to include sufficient Fe$^+$ ions to create an observable
absorption line.
The thickness of the \ion{H}{2} region increases with ionization 
parameter, a fact that explains the strong correlation between the
column density and the ionization parameter.   The net mass outflow
rate is a function of the velocity, explaining the strong correlation
between those two parameters.  The force multiplier depends inversely
on the ionization parameter.  

We also correlated the {\it SimBAL} parameters for the $E1$-divided
samples (Fig.~\ref{correlation_simbal}).  As noted above, $E1$ is
related to the \citet{bg92} Eigenvector 1, which dominates the
variance in quasar properties.  The motivation for looking for
correlations among the $E1$-divided samples is that by removing that
dominant dependence, more subtle parameter 
dependencies may be revealed.

\begin{figure*}[!t]
\epsscale{1.0}
\begin{center}
\includegraphics[width=2.75in]{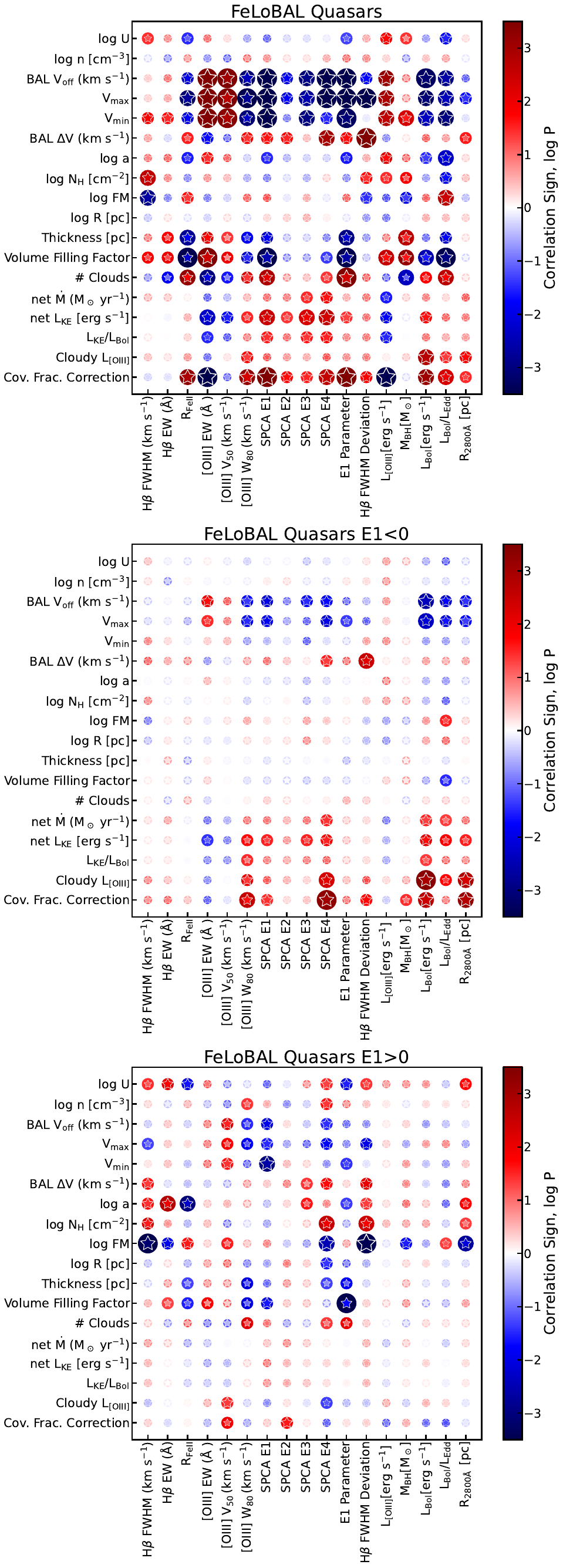}
\caption{The results of the Spearman rank correlation analysis for the
  17 optical emission line parameters and global properties and the 18  
  {\it SimBAL} absorption-line and derived properties.    The
  symbols have the same meaning as in 
  Fig.~\ref{correlation_simbal}.  The top  plot shows the results for
  30-object low-redshift sample, while the middle
  and bottom plots show the   results for the  $E1<0$ and $E1>0$
  subsamples   respectively.  Among the optical parameters, $E1$ and
  related   parameters $R_\mathrm{FeII}$, [\ion{O}{3}] equivalent
  width, SPCA1,  $E1$, $L_\mathrm{Bol}$  and
  $L_\mathrm{Bol}/L_\mathrm{Edd}$, are   most   strongly 
  correlated with the   {\it SimBAL} parameters.   Among the {\it
    SimBAL} parameters, the offset velocities, and  parameters
  associated with the volume filling fraction are the most strongly
  correlated with the   optical   parameters.  \label{corr_optical_simbal}}   
\end{center}
\end{figure*}

Finally, Fig.~\ref{corr_optical_simbal} shows the correlations between
the seventeen optical and global parameters and the eighteen {\it
  SimBAL} parameters.     A special technique was used to handle  
  these data because there are thirty objects and 36 outflow
  components.   Five of the objects showed multiple outflow
components\footnote{SDSS~J025858.17$-$002827.0,
  SDSS~J103903.03$+$395445.8, SDSS~J104459.60$+$365605.1,
  SDSS~J112526.12+002901.3, and SDSS~J144800.15+404311.7.
  SDSS~J144800.15+404311.7 has three outflow components; the other
  objects have two.}; \citet[][\S6.2,]{choi22} discussed the
need for multiple  outflow  components in these objects.  
 We assumed that physically
  one of the   components is   more representative than the other one
  or two.  For   example,   [\ion{O}{3}] may be produced by one
  component in the outflow but not   another.  Likewise, the
  outflow location may be correlated with the outflow velocity for
  components that share some fundamental property (i.e., perhaps they
  are the main outflow in the system), but other, subsidiary outflows
  do not obey this trend.   Therefore, we
  computed correlations among all   combinations and present the
  statistically most significant one.    

\section{Analysis}\label{analysis}

\subsection{The Location of the Outflow}\label{logr}

\begin{figure*}[!t]
\epsscale{1.0}
\begin{center}
\includegraphics[width=6.5truein]{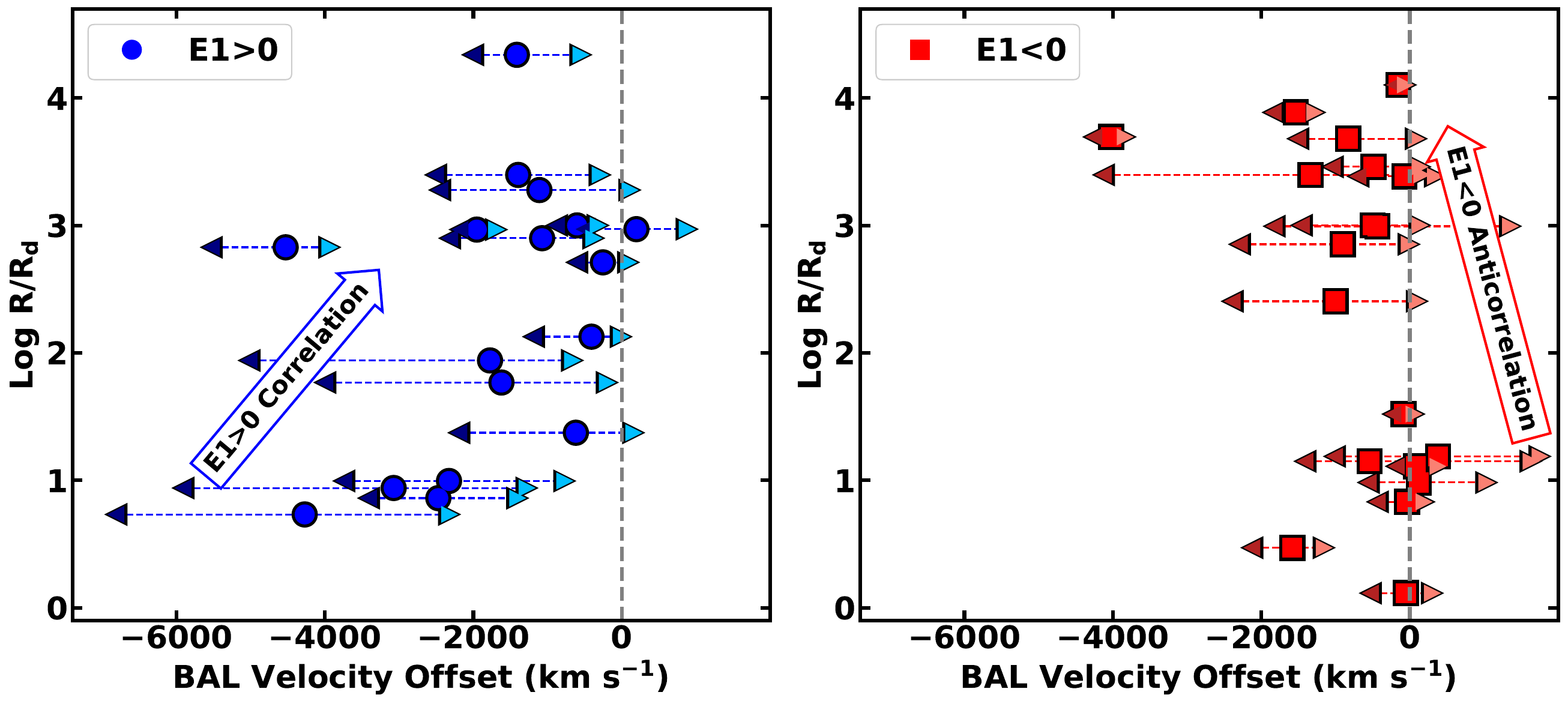}
\caption{The location of the outflow determined using
  {\it SimBAL} modeling \citep{choi22} and normalized by the dust
  sublimation radius 
  as a function of the outflow velocity.  Negative velocities denote
  outflows.  The circles and squares show 
  the column-density weighted velocity, while the left and right
  triangles show the maximum and minimum speeds, respectively.  There
  is an anticorrelation (correlation) between $\log R/R_\mathrm{d}$
  and the outflow velocity for the $E1 <0$ ($E1>0$) FeLoBALQs.
  These different behaviors imply a  difference in the formation
  and acceleration of outflows for the two   classes of FeLoBAL
  quasars.    \label{logr_voffset}}        
\end{center}
\end{figure*}

An interesting set of patterns observed among the {\it SimBAL} 
parameters is shown in Fig.~\ref{logr_voffset}.   We plot the velocity
of the outflow as a function of the log of the radius divided by the
dust sublimation radius \citep[$R_d=0.16  L_{45}^{1/2}\rm \,
  pc$,][]{en16}; note that $R_\mathrm{d}$ is close to 1 parsec in
these samples.  It is immediately apparent that among the compact
outflows $\log R/R_d < \sim 2$, the $E1>0$ objects have
systematically larger velocity outflows than the $E1<0$ objects. 

While there is no correlation between the location of the outflow
$\log R/R_\mathrm{d}$ and the outflow velocity for the sample as a
whole, we found a tentative or marginal correlation for the $E1>0$
subsample and an anticorrelation for the $E1<0$ subsample (p-values are reported later in the text).
$E1<0$ objects with outflows close to the central engine generally have no
net outflow velocity, while those located far from the central engine
show modest outflows.  In contrast, some of the largest velocities in
the sample are found in $E1>0$ object outflows located close to the
central engine, while  at larger distances, the velocities tend to
be lower.  

The correlation between the outflow velocity and $\log R/R_\mathrm{d}$
observed among the $E1>0$ objects might  
be expected from a high accretion rate quasar.  The terminal velocity
for a radiatively driven outflow is larger at smaller radii because of
the higher available photon momentum, which
means that larger velocity at small radii would be needed for a
sustained (i.e., not failed) wind.   

\citet{choi22} identified a new class of FeLoBAL quasars
called ``loitering'' outflows as objects that have $\log R < 1$ [pc] and
velocity offset of the excited state \ion{Fe}{2}$\lambda 2757$
$v_\mathrm{off,FeII excited} < 2000\rm \, km\, s^{-1}$
\citep[Fig.\ 18][]{choi22}.  They  
found that the loitering outflow objects had distinct photoionization
properties too: they have larger ionization parameters 
and densities compared with the full sample. 
\citet{leighly22} found that in the low-redshift subsample,
almost all loitering outflow objects had $E1<0$ and were
therefore categorized as low accretion rate objects.  

We next try to understand the origin of the low velocities among
the loitering outflows.  The high ionization parameter 
\citep[$-2 < \log U < 0.5$, Fig.\ 6][]{choi22} and accompanying
large column density 
\citep[$\log  N_H > 21.5$ (cm$^{-2}$), Fig.\ 7][]{choi22}
yield a low force multiplier;  basically, the   slab has very large
column of   gas where the illuminated face and a   significant
fraction of the  total column density is too ionized to   contribute
much to   resonance  scattering \citep[e.g.,][]{arav94b}.  The
$E1<0$ objects are characterized by a low Eddington ratio 
($\log  L_\mathrm{Bol}/L_\mathrm{Edd} < -0.5$), which means that the
radiative flux is small relative to the gravitational binding of the
black hole and therefore it is less able to accelerate the outflow
gas. Thus, the combination of the large outflow column and low
radiative flux compared with gravity  may explain the low velocities 
\citep[$V_\mathrm{off} < -2000\rm \, km\, s^{-1}$,
    Fig.\ 18][]{choi22}. 

However, not all $E1<0$ objects have loitering outflows; at larger
radii, near $\log R/R_\mathrm{d}=3$, the $E1<0$ objects merge with the
$E1>0$ objects, and outflows of both $E1$ groups have velocities near
$-1000\rm \, km\, s^{-1}$.   The combination of near-zero velocity 
for small $\log R/R_\mathrm{d} < 2$ and outflows for larger
$\log   R/R_\mathrm{d} > 2$ results in the anticorrelation
between the outflow velocity and $\log R/R_\mathrm{d}$ among the
$E1<0$ objects shown in Fig.~\ref{logr_voffset}. 

As seen in  Fig.~\ref{simbal_dist}, it is clear that the  $E1>0$
objects have systematically larger velocity outflows than the $E1<0$
objects; the probability that they are drawn from the same population
is less than 1.6\% (Table~\ref{tab_distributions}).  However, 
correlations between the outflow velocity and the location of the
outflow are barely or arguably not significant, and might be construed
to fall in the realm of p-hacking.  For example, if we examine $\log
R$ as a function of V$_{off}$, then the $E1<0$ correlation is significant
($p=0.025$) but the $E1>0$ is not ($p=0.076$), while for  $\log
R/R_d$, the $E1<0$ correlation is not significant ($p=0.052$) while
$E1>0$ is significant ($p=0.050$).  On the other hand, these
correlations are principally driven by the differences in the velocity
distributions at $\log R < 2$ [pc], which are very clear from
Fig.~\ref{logr_voffset}.  Regardless, these differences in the
behavior in outflow velocity among the $E1<0$ and $E1>0$ objects may
point to a difference in formation and acceleration of outflows among
those two classes of objects.  In particular, if the $E1$ parameter is 
considered to be a proxy for $L_\mathrm{Bol}/L_\mathrm{Edd}$, and
these differing behaviors may point to a difference in formation and
acceleration of outflows as a function of that parameter.
The relationship between potential acceleration mechanisms and the
  origin and location of FeLoBAL winds is discussed extensively in
  \citet{choi22} \S7.2.  

\begin{figure*}[!t]
\epsscale{1.0}
\begin{center}
\includegraphics[width=6.5truein]{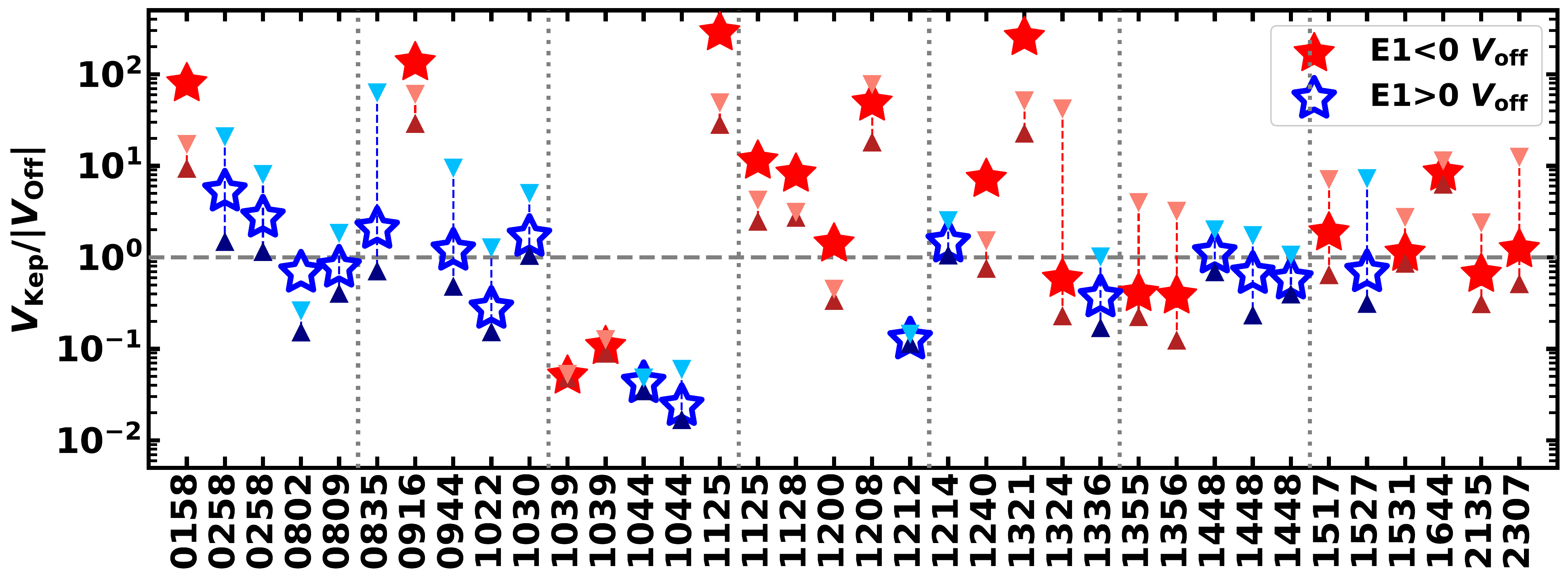}
\caption{The ratio of the Keplerian velocity at the location of the
  outflow to the absolute value of the outflow velocities.  The stars
  denote $V_\mathrm{off}$.  The triangles show $V_\mathrm{min}$
  and $V_\mathrm{max}$, and can be offset from the stars if the
  trough includes both inflow and outflow.  Very large values of this
  ratio show absorption  features that have velocities much lower
  than the Keplerian velocity.      \label{vkep_rat}}        
\end{center}
\end{figure*}

Using the black hole masses estimated in \citet{leighly22}
we computed the Keplerian velocities at the location of the outflow.
We then computed the ratio of the Keplerian 
velocity to the outflow velocity $V_\mathrm{off}$, as well as the
ratio to $V_\mathrm{min}$ and $V_\mathrm{max}$, and plot the results
in Fig.~\ref{vkep_rat}.  This plot shows 
several interesting features.   Large values of
$V_\mathrm{Kep}/\lvert V_\mathrm{off} \rvert$   indicate absorption features
that have line-of-sight velocities 
much lower than the local Keplerian velocity (e.g., SDSS~J1125+0029,
SDSS~J1321+5617).  A number of $E1<0$ objects fall into this
category.  Because the continuum emission region is so small
compared with the outflow location (\S\ref{volumn_filling}), this
phenomenon could be achieved if the outflow velocity vector lies
strictly along the line of sight to the nucleus, and the local
Keplerian motion is principally tangential.  

On the other hand, very low values of
$V_\mathrm{Kep}/\lvert V_\mathrm{off} \rvert$  indicate objects with
outflow velocities very much larger than the local Keplerian velocity
(e.g., SDSS~J1039+3954, SDSS~J1044+3656).   Small values 
suggest that the outflow is kinematically decoupled from the
gravitational potential of the black hole.  It is possible that these
small ratios signal a distinct acceleration mechanism for their
outflows, such as the ``cloud-crushing'' mechanism suggested by
\citet{fg12}, rather than radiative line driving or dust acceleration,
where the magnitude of those mechanisms scales with the location of the
outflow.  

Finally, a number of the outflows show
$V_\mathrm{Kep}/\lvert V_\mathrm{off} \rvert$    commensurate with unity.
Most of the $E1>0$ objects fall into this category.  Of particular
interest is SDSS~J1448+4043.   \citet{choi22} found that the
{\it SimBAL} solution required three separate outflow components in
this object (\S~6.2 of that paper). The lowest velocity component
could be seen to be kinematically distinct, since it is narrow and
shows a prominent ground-state \ion{Mg}{1}$\lambda 2853$ line.  The
other two features are kinematically blended in this
overlapping-trough object and were inferred to be distinct based on
their photoionization properties \citep[Fig.\ 11][]{choi22}. They
found that the outflows lie at dramatically different distances ($\log
R=0.84$, 2.05, and $>3.1$ 
[pc]), yet their outflow velocities are commensurate with Keplerian
velocity at those locations.  These outflows could be
considered kinematically coupled in some way to the gravitational
potential of the black hole.  

\subsection{Bolometric Luminosity}\label{lbol}

\begin{figure*}[!t]
\epsscale{1.0}
\begin{center}
\includegraphics[width=6.5truein]{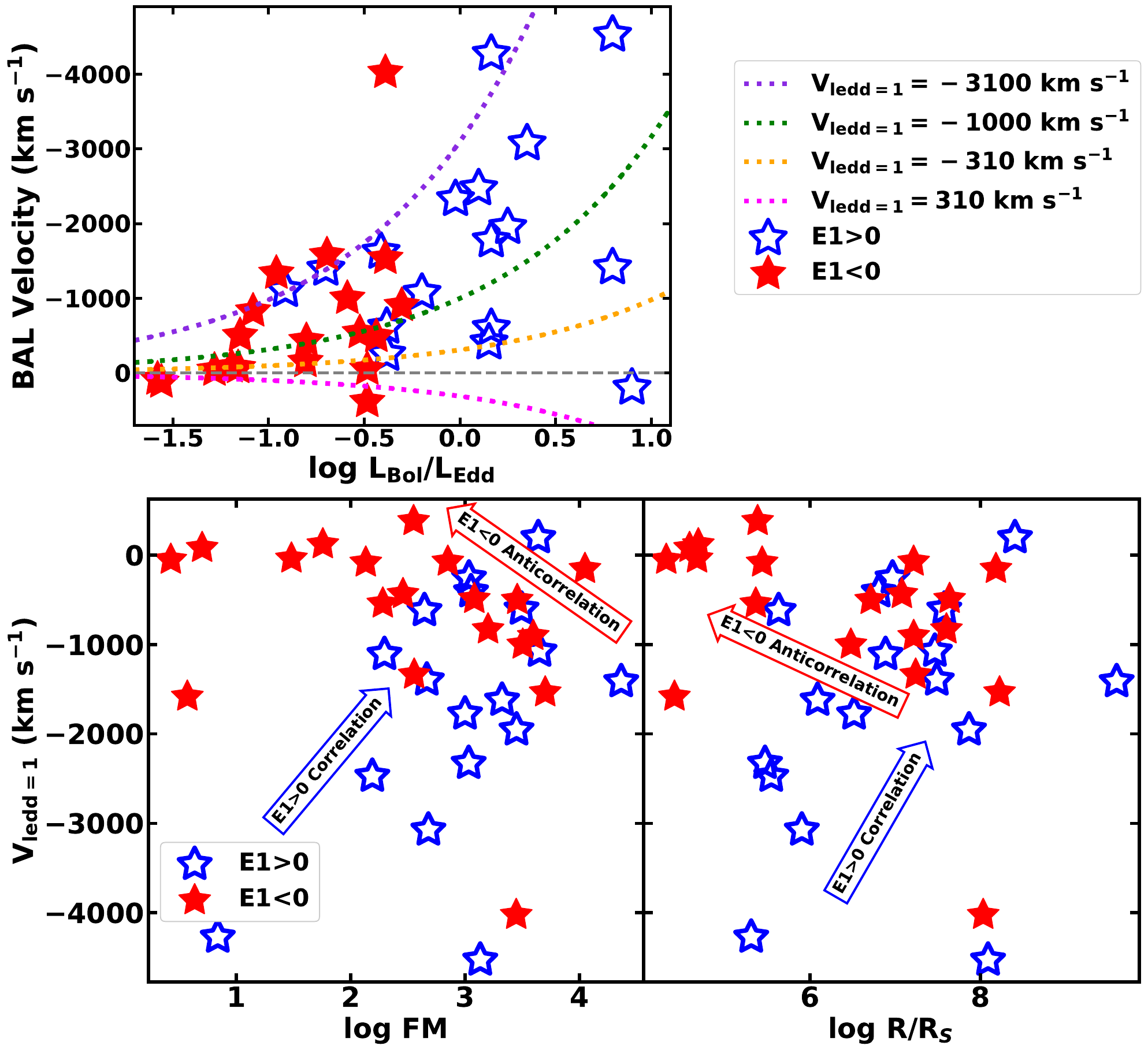}
\caption{The BAL outflow velocity as a function of Eddington ratio for
  the sample of low-redshift  FeLoBAL quasars. {\it Top:} The sample
  shows the $V\propto L_\mathrm{bol}/L_\mathrm{Edd}$ upper-limit
  envelope behavior commonly found in BALQs \citep{laor02,
      ganguly07}; see text for details. The scatter below the 
  envelope can be characterized by the parameter $V_\mathrm{ledd=1}$, the velocity
  any object would have if $L_\mathrm{bol}/L_\mathrm{Edd}=1$ (see
  text for   details).    {\it Bottom Left:} An anticorrelation
  between the force   multiplier and  $V_\mathrm{ledd=1}$
  indicates that the scatter in the $E1<0$ objects may be caused by
  the scatter in $FM$.  {\it Bottom Right:}  A correlation of $R/R_S$ with $V_\mathrm{ledd=1}$
  suggests that the scatter in the  $E1>0$ objects may be caused by
  the scatter in   $R/R_S$.  \label{lum_vel}}         
\end{center}
\end{figure*}

The plot illustrating the results of the correlation analysis
(Fig.~\ref{corr_optical_simbal}) reveals the most 
frequently observed relationship  among BAL quasar outflows:
the correlation of the luminosity and Eddington ratio with the outflow
speed ($p=5\times 10^{-4}$ and $p=1.2\times 10^{-3}$, respectively).
The relationship between the BAL outflow velocity and 
Eddington ratio is shown in Fig.~\ref{lum_vel}.  These relationships
have been previously   reported for HiBAL quasars \citep{laor02,
  ganguly07,gibson09}, and are  generally found among objects with
outflows \citep[e.g.,][]{fiore17}. \citet{ganguly07} noted that the
terminal velocity of an outflow should scale with Eddington ratio as
$v_\mathrm{terminal} \propto (L_\mathrm{bol}/L_\mathrm{Edd})^{1/2}$
\citep{hamann98a, misawa07}; this dependence arises from the solution
of the conservation of momentum equation \citep[e.g.,][]{leighly09}
assuming that the outflow is accelerated by   
radiation. Thus the outflow velocity is predicted to be
correlated with the luminosity relative to the Eddington value.
However, a linear relationship between these two quantities is not what
is observed; rather, there is usually an upper-limit envelope of 
velocity as a function of luminosity or
$L_\mathrm{bol}/L_\mathrm{Edd}$.  That is, at
any luminosity, there is a range of outflow velocities up to some
upper limit value, and those upper limit values are correlated with
luminosity.  For example, see Fig.\ 6 in \citet{laor02} and Figs.\ 6
and 7 in \citet{ganguly07}. This upper-limit relationship is seen in
our data too (Fig.~\ref{lum_vel}, top). 

Theoretically, the speed of a radiatively driven outflow should also
depend on two factors: 1.)  the launch radius (because the flux
density of the radiation drops as $R^{-2}$) and 2.) the force
multiplier (because the ability of an outflow to use the photon
momentum depends on which ions are available that can scatter the
photons).  Using the 
momentum conservation equation it can be shown that $$v \sim FM^{1/2}
(L/L_\mathrm{Edd})^{1/2} (R_\mathrm{launch}/R_\mathrm{S})^{-1/2},$$
where $R_\mathrm{S}$ is the Schwarzschild radius.  The launch radius is
difficult to determine, but we can estimate it by assuming that the
measured velocity is some representative fraction of the terminal
velocity.  For a $\beta$ velocity law\footnote{A   $\beta$ velocity
  law is expressed as $v(r)=v_\infty (1-R_\mathrm{Launch}/r)^{\beta}$
  where $v_\infty$ is the terminal velocity and $R_\mathrm{Launch}$
  is the launch radius.  It is   often used   for modeling winds from
  hot stars.}, the launch radius is a
constant fraction of the radius at which the outflow is observed, i.e.,
$R_\mathrm{launch}/R_\mathrm{S}\approx R/R_\mathrm{S}$.  While this
approximation would  be inadequate to solve an equation of motion for a
single object \citep[e.g.,][]{choi20}, it may be sufficient to
compare a sample of objects.  The force multiplier, defined to be the
ratio of the total opacity to the electron scattering opacity, is an
output of {\it Cloudy}; it includes both line and continuum opacity.  

Armed with these relationships and approximation, we can investigate
whether the scatter in outflow velocity for a given 
$L_\mathrm{bol}/L_\mathrm{Edd}$  could be a consequence of intrinsic
scatter, whether there are trends in either the launch radius or the
force multiplier, or whether both cases may apply. To put it another
way, if the velocity upper-limit envelope describes the behavior of an
optimal outflow, perhaps the objects below that optimal level may be
deficient in either their force multiplier or their launch radius.

We note that there is another factor that could be important: the
angle of the line of sight to the velocity vector.  We do not discuss
that factor here, which means that there should be additional
intrinsic scatter associated with that parameter.  In addition, dust
scattering may be important in accelerating quasar outflows
\citep[e.g.,][]{thompson15, ishibashi17}, and our use of the force
multiplier means that we are only considered continuum and resonance
scattering for the acceleration mechanism. Additional discussion
  of the potential role of dust scattering in acceleration in FeLoBAL
  quasar outflows can be found in \citet{choi22} \S7.2.

We investigated these questions by first defining a parameter $V_\mathrm{ledd=1}$
that describes how far below the optimal outflow velocity an object
lies on the Eddington ratio versus velocity plot shown in
Fig.~\ref{lum_vel}.   That is, we solved $V_\mathrm{off}=V_\mathrm{ledd=1}
(L_\mathrm{bol}/L_\mathrm{Edd})^{1/2}$ for $V_\mathrm{ledd=1}$, where
$V_\mathrm{off}$ is plotted as the $y$ axis in the top panel of
Fig.~\ref{lum_vel}.   Then, for any
particular object, $V_\mathrm{ledd=1}$ is the velocity that it would have if
$L_\mathrm{bol}/L_\mathrm{Edd}=1$.  In essence, we have derived a
parameter that can be used to compare objects as though they have the
same $L_\mathrm{bol}/L_\mathrm{Edd}$ (similar to the concept of absolute
magnitude).   Traces for representative values
of $V_\mathrm{ledd=1}$ are shown in Fig.~\ref{lum_vel}.  We then
plotted  this
parameter against the force multiplier and $R/R_S$ (Fig.~\ref{lum_vel}).

We first considered the force multiplier.  We found that the force
multiplier is not correlated with $V_\mathrm{ledd=1}$ for the whole sample, but it
is marginally anticorrelated for the $E1<0$ objects ($r_s=-0.46$, $p=0.047$), and
correlated for the $E1>0$ objects ($r_s=0.62$, $p=6.7\times
10^{-3}$).   Keeping in mind that $V_\mathrm{ledd=1}$ is 
negative for outflows, this means that $E1<0$ objects with larger $FM$
have higher outflow velocities.  We interpret this behavior to mean
that unfavorable $FM$ (too small) is responsible for non-optimal (i.e.,
below the envelope) outflows in the $E1<0$ objects.  In contrast, the
correlation for the $E1>0$ objects means that the $E1>0$ objects with
low $FM$ have higher outflow velocities, opposite of the physical
expectation.  We interpret this to mean that $E1>0$ objects produce
their outflows {\it   despite} non-optimal $FM$ values, and therefore
another factor is causing the below-envelope scatter in the $E1>0$
objects.   

Turning to $R/R_S$, we found that, while there is no correlation between
this parameter and $V_\mathrm{ledd=1}$ for the sample as a whole, there is an
anticorrelation for the $E1<0$ objects ($r_s=-0.52$, $p=0.023$), and a
correlation for the $E1>0$ objects ($r_s=0.51$,
$p=0.035$).  Again keeping in mind that $V_\mathrm{ledd=1}$ is negative for
outflows, the correlation between $R/R_S$ and $V_\mathrm{ledd=1}$ means that objects
with smaller $R/R_S$ have larger outflow speeds, as predicted by
the solution to the momentum conservation equation above.  This
correlation suggests that the below-envelope scatter among the $E1>0$
objects is caused by unfavorable (too large) $R/R_S$
(Fig.~\ref{lum_vel}, bottom panel).  The anticorrelation seen among
the $E1<0$ objects can be interpreted as implying that $R/R_S$ is less
important in determining their outflow velocity.  

Finally, we can compare the $V_\mathrm{ledd=1}$ values for the $E1<0$
and the $E1>0$ objects.  It turns out that the $V_\mathrm{ledd=1}$
distributions for the two classes are statistically indistinguishable.
That is, both types of  objects lie along the same set of
$V_\mathrm{ledd=1}$ traces in Fig.~\ref{lum_vel}. The $E1<0$ objects
have both systematically lower outflow  velocities
(Fig.~\ref{simbal_dist})  and systematically lower
$L_\mathrm{bol}/L_\mathrm{Edd}$ \citep[Fig.~7,][]{leighly22}.
This 
result implies that $E1>0$ objects reach a larger velocity because of
their larger $L_\mathrm{bol}/L_\mathrm{Edd}$, i.e.,
$L_\mathrm{bol}/L_\mathrm{Edd}$ is  primary for both classes of
objects.   

We have shown that in the low-redshift sub-sample of FeLoBAL quasars
the outflow velocity depends on the $L_\mathrm{bol}/L_\mathrm{Edd}$.
This is not a new result; it has been seen before in other samples 
\citep[e.g.,][]{ganguly07}.  Such a result is relatively simple to
extract from any set of BAL quasars, depending as it does  only on
estimation of the outflow velocity (for example, from the \ion{C}{4}
trough), and an estimate of the Eddington luminosity.  That requires
an estimate of the black hole mass, which is arguably most reliably
extracted from H$\beta$ but can also be estimated using \ion{Mg}{2}
and \ion{C}{4}, in principal, although difficult in FeLoBAL quasars
due to the heavy absorption.  The difference in our analysis is that
because {\it SimBAL} yields the physical conditions of the outflow
(including $\log U$, $\log n$, $\log N_{H}$) we can investigate this
relationship in more detail and parse the dependence on the subsidiary
parameters: force  multiplier and estimated launch radius.  We
discovered a difference in velocity dependence on these two parameters
among the two accretion classes.  So while
$L_\mathrm{bol}/L_\mathrm{Edd}$ is principally responsible for
determining the outflow velocity, the force multiplier (launch radius)
is responsible for producing the scatter in the velocity at a
particular value of $L_\mathrm{bol}/L_\mathrm{Edd}$ in the $E1 <0$
($E1>0$) objects.  This result provides additional evidence for
differences in acceleration mechanism.  

\subsection{Geometry Properties of the Outflow}\label{volumn_filling}

\subsubsection{The Full Sample}\label{full}

The volume filling factor ($\Delta R/R,\ \Delta R=N_H/n_H$) gives us
information about the physical size scales of the outflowing gas.  It
is most directly interpreted as the fractional volume of space
occupied by the outflow.  Typically, using the values for column
  density, density, and radius derived using excited state absorption
  lines,  small log volume filling factors, mostly ranging
between $-6$   to $-4$ are found
\citep[e.g.,][]{korista08,moe09,dunn10}.  The volume occupied by  the
absorbing clouds ranges from 0.01\% to 1\%. 
The volume filling 
factor tells us how thin or extended in the radial
direction the BAL cloud structure is and provides us with
information about the BAL physical conditions.  A small volume
filling factor ($\log \Delta R/R\sim -5$) for BALs may imply a
pancake- or shell-like geometry that is very thin in the radial
direction \citep[e.g.,][]{gabel06,hamann11,hamann13}.  These BAL
absorbers with $\log \Delta R/R\lesssim-3$ may be composed of smaller
gas clouds \citep[e.g.,][]{waters19} that are potentially supported by
magnetic confinement in order to avoid dissipation
\citep[e.g.,][]{dekool95}. In contrast, \citet{murray97}  proposed
that a continuous flow from the accretion disk  is the origin of broad
emission lines and BAL  
features; such a flow would have a volume filling factor of
1.  We emphasize that our results are not consistent with
a direct observation of a disk wind because the size scales that we
measure are too large. The minimum distance of the outflow from the
central engine found in our sample is $R\sim 1$ pc in SDSS~J1125$+$0029,
whereas reasonable size scales for disk wind outflows should be
comparable to the size of the accretion disk ($R\ll 0.01$ pc). 
That does not imply that disk winds do not exist but rather that we do
not find them to have rest-UV BAL outflow signatures.
This result is consistent with the literature; among the FeLoBAL quasars
  previously subjected to detailed analysis, typical outflow distances
  lie between 0.4 and 700 parsecs \citep[e.g.,][]{dekool01, dekool02a,
    dekool02b, moe09, dunn10, aoki11, lucy14, shi16, hamann19b,
    choi20}, i.e., no closer than the broad line region.

\begin{figure*}[t]
\includegraphics[width=.48\linewidth]{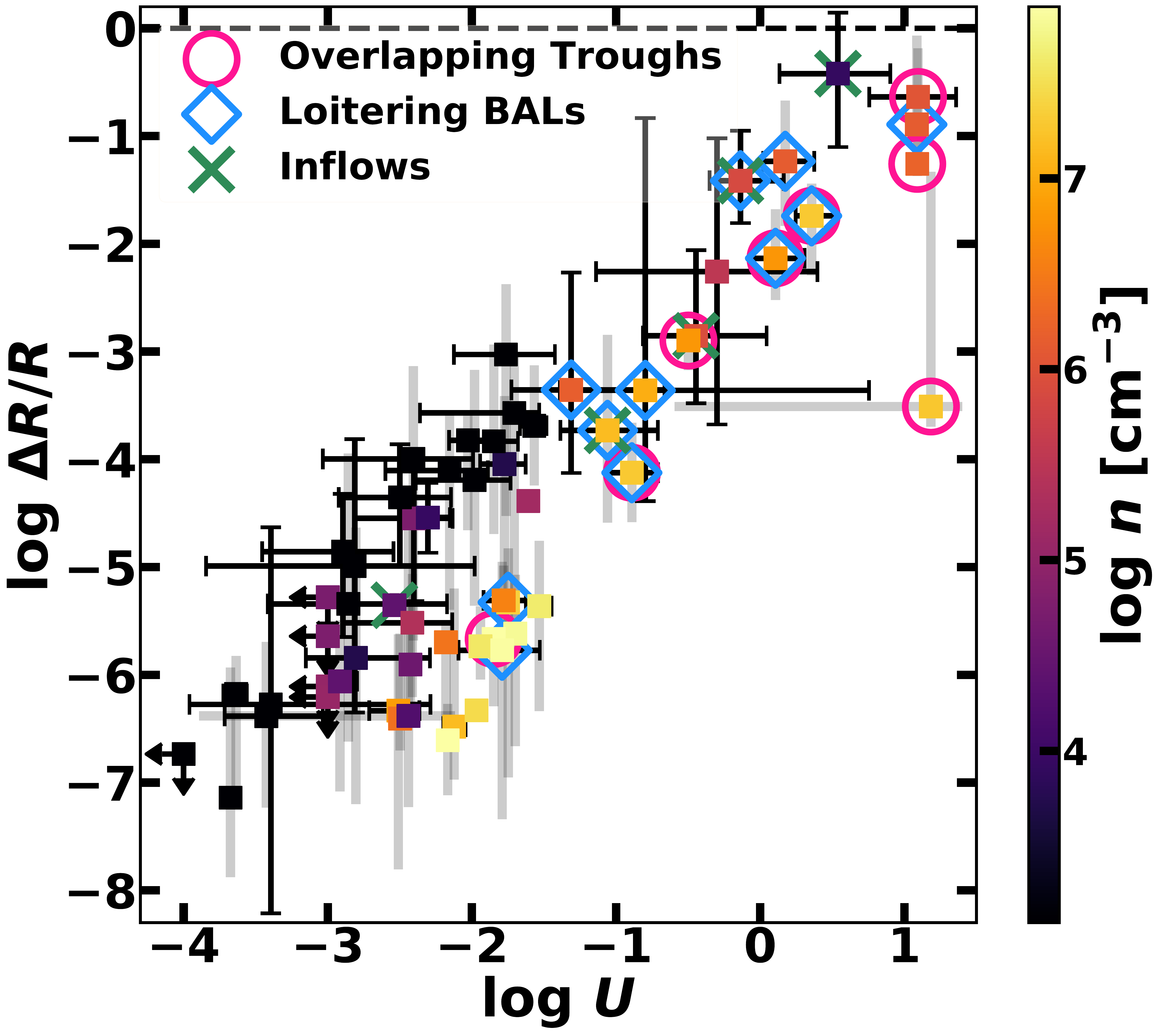}
\includegraphics[width=.51\linewidth]{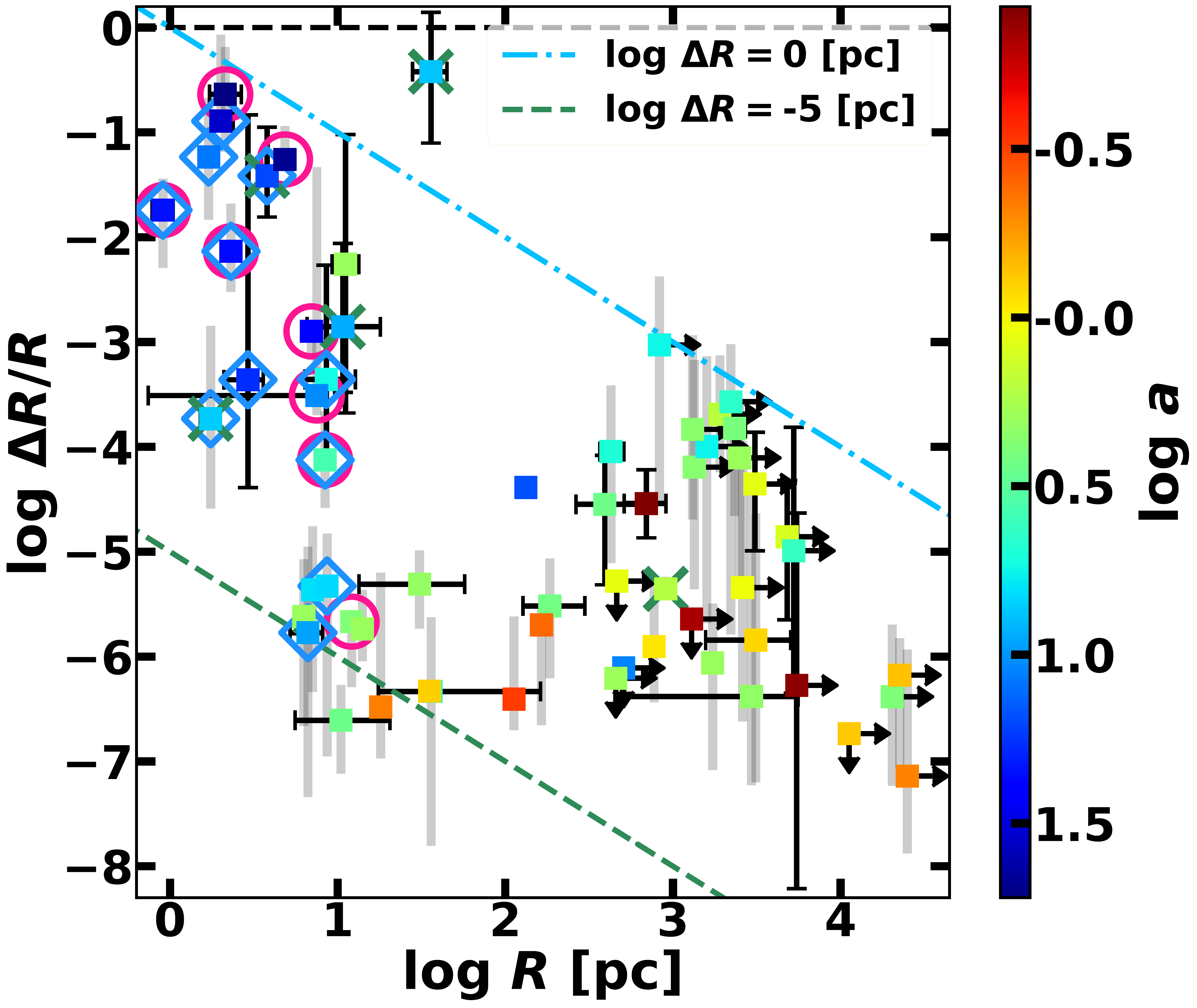}
\caption{The volume filling factor ($\log \Delta R/R$) as a function
  of the ionization parameter ($\log U$) and the location of the
  outflow ($\log R$).  We found a wide range of $\log \Delta R/R$,
  with the overlapping trough and loitering BALs having higher values
  of $\log \Delta R/R$.  {\it Left panel}: $\log \Delta R/R$ increases
  with $\log U$ following the slope of $\sim1.5$.  This tight
  correlation is expected given the relationship between $\log U$ and
  $\log N_H$ for FeLoBALs \citep[\S~5.3,][]{choi22}.  One of the
  main sources of the scatter along the $\log \Delta R/R$ can be
  ascribed to the range of $\log n$ observed in the sample.  {\it
    Right panel}: We found a wider range of $\log \Delta R/R$
  distribution for the FeLoBALs that are located close to the central
  black holes.  The green (dashed) and blue (dotted-dashed) diagonal
  lines represent the locations of the constant physical thickness of
  the BAL winds at $\log \Delta R=-5$ and 0 [pc], respectively.  The
  error bars show 2$\sigma$ uncertainties and the grey shaded bars
  represent the range of the values among the tophat model bins for
  each BAL.
\label{fig:logdRR}}
\end{figure*}

We calculated the volume filling factors for the full sample and
examined the dependence on BAL properties (Fig.~\ref{fig:logdRR}). The
strong correlation seen between $\log \Delta R/R$ and $\log U$ can be
explained by the mathematical relationship between the parameters as
follows. First, the BAL physical thickness ($\Delta R$) is
proportional to the hydrogen column density ($N_H$) which is also
proportional to the ionization parameter $U$ since $\log N_H-\log U$
is nearly constant in the FeLoBALQ sample, and the distance of the
outflow from the central SMBH ($R$) is inversely proportional to
$U^{1/2}$, both for a fixed density.  Dividing the BAL thickness by
its distance from the center, we obtain the volume filling factor
$\Delta R/R\propto U^{1.5}$, and we find a slope of $\sim1.5$ in the
left panel of Figure~\ref{fig:logdRR}.  We observe a scatter around
that line because of the range of $\log N_H-\log U$ and $\log n$ for
the FeLoBALs in our sample  \citep[Fig.~4,][]{choi22}.  There
is also a range in photoionizing flux $Q$ which we assume to be
proportional to $L_\mathrm{Bol}$.  This parameter enters through
$U=Q/4\pi R^2 nc$. Therefore, larger values of $\log N_H-\log U$
(thicker outflows), smaller density, or smaller $\log L_{bol}$
correspond to a larger value of  $\log \Delta R/R$.

The distribution of $\log \Delta R/R$ is not uniform
across $\log R$. At large radii, corresponding to $\log U\lesssim-1$,
the volume filling factors  mostly range between $-6$ to $-4$.  These
values are similar to those reported in the literature for samples of
high-ionization BAL quasars
\citep[e.g.,][]{gabel06,hamann11,hamann13}.  In contrast, the outflows
that are found at $\log R\lesssim1$ have a  very wide range of $\log
\Delta R/R$, ranging from $-6$ to nearly almost zero.  These are
mostly the special types of BALs that were identified in
\citet{choi22}, including the overlapping trough and loitering BALs.      

The analysis for the full sample shows significant differences in
$\log \Delta R/R$ as a function of  radius.  This result suggests that
BAL winds may favor different models at different radii
\citep[][Leighly et al.\ in prep.]{choi22}.   The compact
winds at $\log R\lesssim1$ [pc] showed a wide range of $\log \Delta
R/R$ that agrees with the predictions of the various BAL physical
models that explain either thin shell-like outflows (small volume
filling factor) or stream-like outflows (large volume filling factor).
On the other hand, the properties of distant BAL winds only favor the
physical model with thin pancake-like BAL geometry.  Specifically,
\citet{fg12} proposed that FeLoBALs with large $\log R\gtrsim3$ [pc]
and small $\log \Delta R/R\sim -5$ are formed by ``cloud crushing''
where the ambient ISM is shocked by the supersonic energy-conserving
quasar outflow and the FeLoBALs are formed in-situ at kiloparsec
scales rather than formed near the accretion disk.  In addition, in
order for distant BALs to have large filling factors, the BAL clouds
would need to have large physical radial thicknesses proportional to
their distances from the central engine ($\Delta R\gtrsim10$
pc). Maintaining such large structure is physically challenging due to
cloud destructive processes \citep[e.g.,][]{proga15}. 

\subsubsection{The $z<1$ FeLoBALQs and the {\it E1} Dependence}\label{covfrac}

In this section, we discuss the relationship between the BAL
outflow parameters involving the geometry of the outflow and the
optical-band emission line properties.  We have emission-line
properties only for the 30-object $z<1$ subsample, so this
discussion only involves that subsample.  In particular, we
investigated how the parameters that describe the geometrical
properties of the outflow dependon the $E1$ parameter, and by
extention, the accretion rate.

We first investigated the physical thickness of the gas $\Delta R$.
We showed in \S~\ref{distributions} that the $\Delta R$ parameter is
significantly different for the $E1<0$ and $E1>0$ subsamples
(Fig.~\ref{simbal_dist} and Table~\ref{tab_distributions}); the median
$\log$ thickness is about 1 dex larger for the $E1<0$ objects.  This
result arises because although there is no statistical difference in
$\log U$ between the $E1<0$ and $E1>0$, there is a tendency for $E1<0$
objects to have larger $\log U$ and therefore thicker outflows.  The
thickness is also anticorrelated with both $R_\mathrm{FeII}$ and the
E1 parameter (Fig.~\ref{corr_optical_simbal}) for the same reason.

We next considered the volume filling factor for the low-redshift
subsample.  We found that $E1<0$ objects have significantly
larger volume filling factors than $E1>0$ objects because
of the significant difference in thickness $\log \Delta R$ but also in
the tendency for $E1<0$ objects to have larger $\log U$ and therefore 
smaller $\log R$, i.e., to be located closer to the central engine.

Because we have black hole mass and accretion rate estimates
\citep{leighly22}, we can estimate the number of spherical clouds 
required to completely cover the continuum emission region.
This
parameter is useful to visualize the BAL absorption region  in
the quasar.  The first ingredient in this computation is the size of
the emission region $R_{2800}$. We calculated the size of the
2800\AA\/ continuum emission region using the procedure described in
\S~6.1 of \citet{leighly19}.  To summarize, we used a simple
sum-of-blackbodies accretion disk model \citep{fkr02} and assigned the
2800\AA\/ radius to be the location where the radially-weighted
brightness dropped by a factor of $e$ from the peak value.  Among the
objects in this sample, this parameter spans a rather small
range of values with 90 percent of the objects having $-2.9 < \log
R_\mathrm{2800} < -2.3$ [pc] \citep[Fig.~8, ][]{leighly22}.
This is likely a consequence of the $T^4$ dependence of the accretion 
disk.

The second consideration is the wide range of angular diameters that
the continuum emission region will subtend at the location of the
outflows.  For example, the continuum emission region will subtend an
angular diameter that is 1000 times larger to a wind located at one
parsec than to one located at 1000 parsecs.  Folding in the small
difference in size of the continuum emission region we found that the
largest angular diameter is presented to the higher velocity 
component of SDSS~J1125+0029 at 18 arc minutes\footnote{For reference,
  the angular diameter of the full   moon is 31 arc minutes.}, and the
smallest is the higher velocity component of SDSS~J1044+3656 at
$4.5\times 10^{-4}$ arc minutes.   

The final ingredient is the transverse size of the absorber.  This
parameter cannot be measured directly from these data since absorption
is a line-of-sight measurement.  Instead, we make the simplifying
order-of-magnitude assumption that the absorbing gas occurs in clouds
and the clouds are approximately spherical.  Making this assumption
yields a transverse size that is equal to $\Delta R$, the thickness of
the absorbing gas. 

Employing these three ingredients (the size of the continuum emission
region $R_\mathrm{2800}$, the angular diameter that continuum emission
region will subtend at the location of the outflow, and the transverse
size of the absorbing cloud), we can determine the number of clouds
required to cover the continuum emission region.  For example, if the
angular size of the cloud from the perspective of an observer located
at the continuum emission region is the same as the angular size of
the continuum emission region at the location of the cloud, then only
one cloud is required to cover the continuum emission
region.\footnote{We note in passing that this is the same concept that
is applied to the measurement of the size of the quasar continuum
emission region using gravitational microlensing. The gravitational
lens caustics of a single star are very small, but if the continuum
emission region has a commensurate angular size (because the star is
located in the quasar host galaxy), its light can be
differentially magnified.} Conversely, if the angular size of the
cloud from the perspective of an observer located at the continuum
emission region is much smaller than the angular size of the continuum
emission region from the perspective of a viewer located at the
outflow, then many clouds are required.  The number of clouds is the
ratio of the area of the continuum emission region and the transverse
area of the cloud.  For this order-of-magnitude computation, we assume
that the continuum emission region is viewed face on.

\begin{figure*}[!t]
\epsscale{1.0}
\begin{center}
\includegraphics[width=4.5truein]{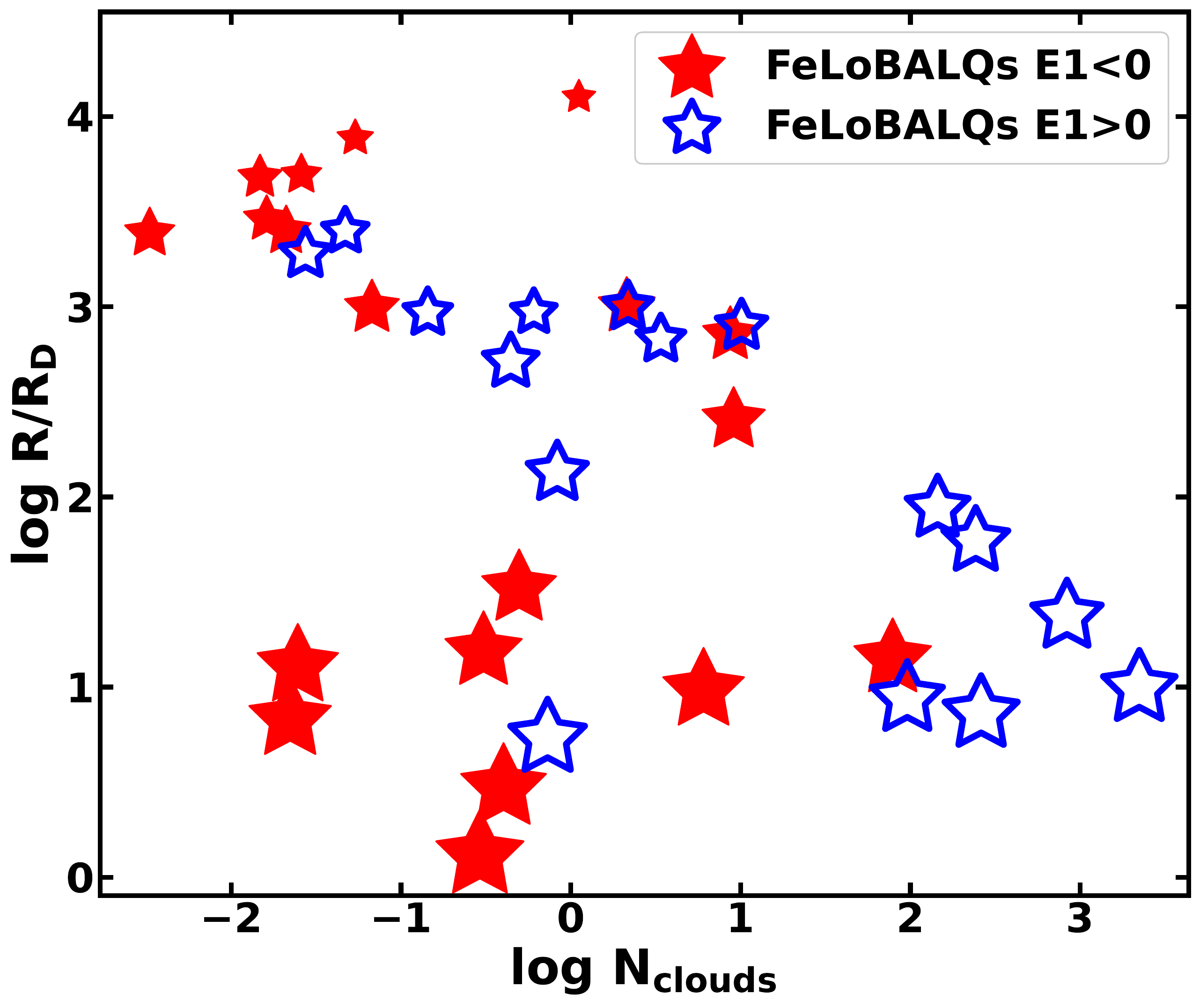}
\caption{Relationship between the number of clouds required to occult
  the 2800\AA\/ continuum emission region and the location of the
  outflow normalized by the dust sublimation radius.  The size of the
  points  is scaled with the angle subtended by the continuum emission
  region as viewed from the location of the outflow. Compact outflows
  (those with   $\log R/R_\mathrm{D} < 2$ are 
  segregated by the   $E1$ parameter, with $E1<0$ objects having
  macroscopic clouds and $E1>0$ objects requiring 100s--1000s of
  cloudlets to cover the source.   \label{numclouds_logr}}   
\end{center}
\end{figure*}

The resulting number of clouds  spans a huge range for the sample.
Ninety percent of the values fall between 0.016 (which means that the
inferred size of the cloud is about 60 times larger than the emission
region) to $\sim 1000$, a factor of more than 60,000.  Moreover, there
is a significant difference between the $E1<0$ and $E1>0$ objects
(Fig.~\ref{simbal_dist}, Table~\ref{tab_distributions}); the median
values of the  log of the number of clouds  are 0.3 and 2.2,
respectively.   

The $E1>0$ outflows located close to the central
engine (lower right in Fig.~\ref{numclouds_logr}) require a large
number of clouds (100s to 1000s) to cover the continuum 
emission region.  Their outflows are characterized by a lower
ionization parameter (typically less than $-1.5$) compared with the
$E1<0$ objects ($\log U\sim 0$), and therefore physical thickness of
the outflow is smaller in these objects.  Physically, this scenario
suggests that the outflow is a fine mist of cloudlets.  

Objects in which the number of clouds is less than 1 are split
between those at large distances from the central engine ($\log
R/R_\mathrm{D} > 2$) and those at small distances ($\log
R/R_\mathrm{D} < 2$).  In these objects, the angular size of the BAL
cloud structure is comparable or larger than the projected angular
size of the continuum emission region.  One possibility is that our
simplifying estimate that the clouds are approximately spherical is
wrong.  For example,  the number of clouds required to cover the
continuum emission region could be much larger if each cloud were long
and needle-like, with the long axis pointed toward the central engine.
This scenario might be somewhat reasonable physically if the clouds
are confined magnetically along field lines that are bent radially by
radiation pressure \citep[e.g.,][]{dekool95} or sheared by radiation
pressure.   

Another possibility is that the nature of the partial covering is
different in some of the objects.  The {\it SimBAL} model includes a
power-law partial covering parameter $\log a$ that may parameterize a
mist of clouds uniformly covering the continuum emission region
\citep[see ][and references therein for discussion and
  visualization]{leighly19}.  The presence of power-law partial
covering does not preclude the presence of step-function partial
covering as well.  The step-function partial covering can be
understood as a partial occultation of the continuum emission region.
There is some evidence for the presence of step-function partial
covering in objects lying in the lower left corner of
Fig.~\ref{numclouds_logr}.  In several of these objects, the {\it
  SimBAL} models required that a portion of the continuum and/or
emission lines be unabsorbed by the outflow \citep[\S6.4,
  Fig.\ 17][]{choi22}.  Physically, this result might be expected when
the continuum emission region has a large angular size from the 
perspective of the absorber, i.e., an absorber with small $\log
R/R_\mathrm{d}$.  It may mean that the outflow is not a mist of clouds
uniformly covering the source, but a distribution of nearly continuous
gas.   

\subsection{[\ion{O}{3}] Emission from the BAL Gas}\label{oiiiemission}

Outflows in quasars are also seen in ionized emission lines.  For
example, kiloparsec-scale outflows observed in emission lines such as
[\ion{O}{3}] are known to be common among luminous AGN
\citep[e.g.,][]{harrison14,bischetti17,vayner21b}.  More compact
ionized emission-line outflows have been resolved in nearby objects;
for example, ionized gas outflows have been found 0.1--3~kpc from the
nucleus \citep{revalski21}.  It is possible that ionized emission-line
outflows and BAL outflows are related. The outflowing broad
absorption line gas is photoionized, and therefore it must produce
line emission.  In particular,  [\ion{O}{3}]$\lambda 5007$ line
emission is an important coolant in photoionized gas
\citep[e.g.,][]{of06}.   It is possible that in some objects the same
gas produces absorption lines along the line of sight as well as
emission lines from all lines of sight.

There are several fundamental problems that make finding a connection
difficult.  While absorption is a line-of-sight effect, an emission
line is an aggregate of many lines of sight, so that emission from gas
not associated with the BAL  would be included in any observed line
emission.  In other words, the covering  fraction of the emitting gas
needs not be the same as that of the  absorbing gas.  [\ion{O}{3}]
emission is observed to have a very large range of equivalent widths
\citep[6--84 
 \AA\/,][]{shen11} potentially originating in a range of gas covering
fractions \citep{bl05b, ludwig09,stern12}.   Moreover, line emissivity
depends on density squared below the critical density, i.e.,
$n_{cr}=6.8\times    10^5\rm \, cm^{-2}$ for  [\ion{O}{3}] 
  $\lambda 5007$], \citep{of06}, and on the density above the critical
density.  Thus, the line emission might not be  seen against the
continuum if the density is too low.  Finally, if the absorption lines are
broad, then the line emission may be distributed over a large range of
velocities, and the line may be too broad to be seen against the
continuum.   

We do not have information about the extent of the line 
emission for our objects.  However, it is interesting to see if there
is a correspondence or relationship between the predicted [\ion{O}{3}]
emission from the BAL gas and the observed [\ion{O}{3}] emission.  
\citet{xu20} tackled this problem using a sample of seven $z\sim 2$
quasars.
Those objects were chosen to have $r$-band magnitude
  $\lesssim 18.8$ and deep \ion{Si}{4}$\lambda\lambda 1393.76, 1402.77$
  troughs. They found evidence for a link between the [\ion{O}{3}]
emission and the BAL absorption.  However, they 
presented a conceptual error: they suggested that [\ion{O}{3}] emission is
suppressed at high densities.  In fact, [\ion{O}{3}] emission always 
increases with density.  Rather than a decrease or suppression of
[\ion{O}{3}] at high densities, the property that decreases is the
ratio of the [\ion{O}{3}] emission with respect to lines with higher
critical densities.  This physics is the basis of the
\ion{Si}{3}]$\lambda 1893$/\ion{C}{3}]$\lambda 1909$ density
    diagnostic used in the near UV    \citep[e.g.,][]{leighly04}.   

We used {\it Cloudy} to predict the [\ion{O}{3}] line emission from the
outflowing gas in each of the 36 BAL components of the thirty $z<1$ 
objects.  Each component is characterized by a single ionization
parameter and density, but is generally split into multiple bins with
different column densities \citep[][]{leighly18,choi22}.  The
[\ion{O}{3}] emission was  computed for 
each bin and then summed.  The local covering
fraction $\log a$ for the BAL outflows was not taken into account in
the computation of the [\ion{O}{3}] flux, since it is not clear how
it would manifest in the observations of line
emission.  The luminosity of the [\ion{O}{3}] emission was then
computed for each component assuming a global covering
fraction of 10\% for the line-emitting gas.

We expect the luminosity of the predicted [\ion{O}{3}] could depend on
several parameters.  
The [\ion{O}{3}] flux density should be larger for higher 
densities.  It should be larger for higher ionization parameters as a
consequence of the larger column density required to include the
hydrogen ionization front in FeLoBAL outflows.  Both of these
conditions are met in outflows closer to the central engine. But the
volume included increases as $R^2$ for the absorbing material for 
a fixed emission-line-region covering fraction. 
Since $R^2\propto \frac{1}{nU}$, the effects cancel out, leaving the
{\it Cloudy} L$_\mathrm{[OIII]}$ uncorrelated with $n$, $U$, or $R$ 
(Fig.~\ref{correlation_simbal}).   

\begin{figure*}[!t]
\epsscale{1.0}
\begin{center}
\includegraphics[width=6.5truein]{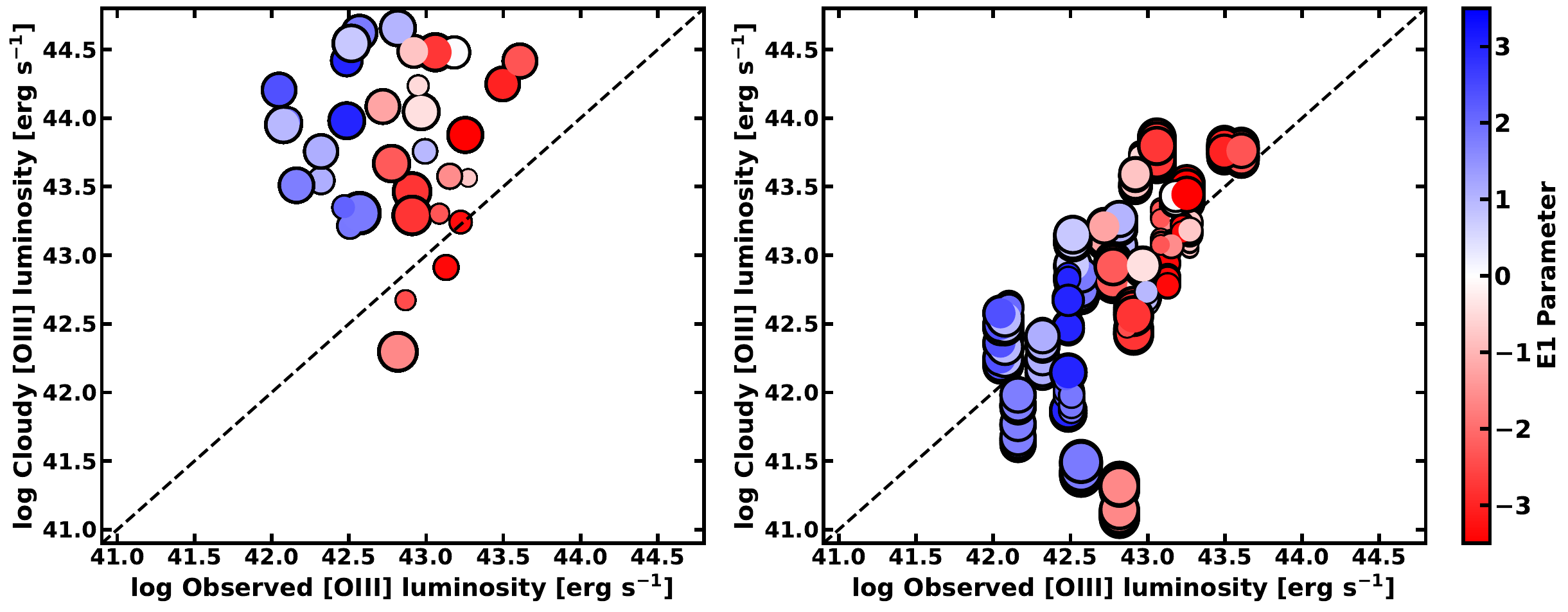}
\caption{The inferred [\ion{O}{3}] luminosity from the {\it SimBAL}
  models  as a function of the
  observed [\ion{O}{3}] luminosity.  The several objects with more
  than one outflow component are plotted more than once. The marker
  color denotes the   corresponding E1 parameter and the size of the
  marker corresponds to   the location $\log R$ parameter.  {\it
    Left:} The predicted values   are generally larger than the
  observed values, potentially implying    a smaller emission-line gas
  covering fraction than the assumed 10\%.   {\it Right:} The same as
  the left plot, with covering fraction   correction determined   by
  the regression analysis (see text).  The   data were plotted using
  regression results for each of the 48    combinations.  \label{lum_vs_lum}}  
\end{center}
\end{figure*}

The left panel of Fig.~\ref{lum_vs_lum} shows the {\it
  Cloudy} predicted [\ion{O}{3}] luminosity as a function of the
observed [\ion{O}{3}] luminosity.  All of the predicted line emission
from objects with multiple BAL outflow components are plotted with
respect to the single observed [\ion{O}{3}]
luminosity.    There is clearly no relationship between these two 
luminosities.  Moreover, while both the observed and predicted
[\ion{O}{3}] luminosities each span about 1.7 dex, the ratio of the
two spans almost 2.7 dex.  In other words, the difference between the
observed and predicted [\ion{O}{3}] emission is not subtle.  

It is possible that the observed and predicted [\ion{O}{3}]
luminosities could be reconciled if the global covering fraction of
outflowing line-emitting gas were not constant for all outflows or for
all emission-line regions.  For example, the location of the outflow in
this sample spans three orders of magnitude, and it is conceivable
that the emission-line covering fraction is different in the vicinity
of the torus ($R\sim 1$ pc) compared with on galaxy scales ($\log R
\sim 3 $  [pc]).  

We parameterized the profound difference between the observed and
predicted [\ion{O}{3}] luminosities by defining the covering fraction
correction factor.  The covering fraction correction factor is the
difference between  the log of the {\it Cloudy} predicted [\ion{O}{3}]
luminosity and the log of the observed [\ion{O}{3}] luminosity.  The
covering fraction correction factor measures how much larger
or smaller than the assumed value of 0.1 that the covering fraction
needs to be to reconcile the observed and predicted [\ion{O}{3}]
luminosities.  In \S~\ref{distributions} we showed that the covering
fraction correction factor is significantly different for $E1<0$
versus $E1>0$ objects.  The median values were 0.6 (1.5) for the $E1<0$
($E1>0$) objects respectively, implying that, on average, the
emission-line region covering fraction needs to be 4 (32) times
smaller than 0.1.  

We explored the 
possibility that the covering fraction depends on other parameters
using a multiple regression analysis.  We considered seven independent 
variables, and our reasons for choosing these parameters follows.  We
included $\log R$ for the reasons outlined 
above.  We also considered Seyfert type using the E1 parameter, and
the log of the estimated bolometric luminosity because there is
evidence for a reduction in [\ion{O}{3}] emission at larger
luminosities \citep{bl05b}.  Objects with a high Eddington ratio might
produce more powerful winds and thereby evacuate a larger fraction of
their reservoir of gas, so we also consider $\log
L_\mathrm{Bol}/L_\mathrm{Edd}$.  
There is evidence that the BAL partial covering fraction $\log a$ is
local \citep{leighly19}, but it is possible there are global trends as
well. In any particular outflow component, $\log a$ is a function of
velocity.  For this experiment, we chose the 
representative $\log a$ to be the one in the bin with the deepest
ground-state \ion{Fe}{2} absorption.  If the velocity of the outflow
is very large, then the resulting line may be very broad and blend
with the continuum, so we also considered the opacity-weighted
$v_{off}$.  Finally, the observed [\ion{O}{3}] velocity offset might
reveal a connection between the outflow and the observed emission.

The result forms a multiple regression problem with the seven
independent variables listed above.  As in \S~\ref{correlations}, we
accounted for the 
multiple components in five of the objects by running the regression
analysis for all 48 combinations.  We used {\tt
  mlinmix\_err} \citep{kelly07} which accounts for measurement errors
in both the dependent variable (the covering fraction) and the
independent variables (the design matrix) using the Pearson
correlation coefficient.  To determine which
independent variables can best reproduce the variance in the covering
fraction, the multiple regression procedure was iterated, each time
removing the independent variable with the largest $p$ value until all
of the remaining variables showed a $p$ value no larger than the cutoff
which was chosen to be 0.025.  

A statistically significant correlation was found between the E1
parameter and the covering fraction correction factor for all 48
combinations.  The next most significant parameter was $\log R$ (39), 
followed by the velocity offset of [\ion{O}{3}] (23).  Since more than
half of the combinations found significant regression with the E1
parameter and $\log R$, we proceeded to extract the regression
parameters for these two parameters and all combinations.   

The right panel of Fig.~\ref{lum_vs_lum} shows the results of the
regression analysis between the covering fraction correction factor
and the independent parameters E1 and $\log R$.  The {\it Cloudy}
covering fraction was adjusted using the best-fitting regression
parameters.   Points are shown for each of the
48 combinations.  The relationship between the observed and {\it
  SimBAL}-predicted [\ion{O}{3}] emission is now linear, although 
considerable scatter remains.  

\begin{figure*}[!t]
\epsscale{1.0}
\begin{center}
\includegraphics[width=6.5truein]{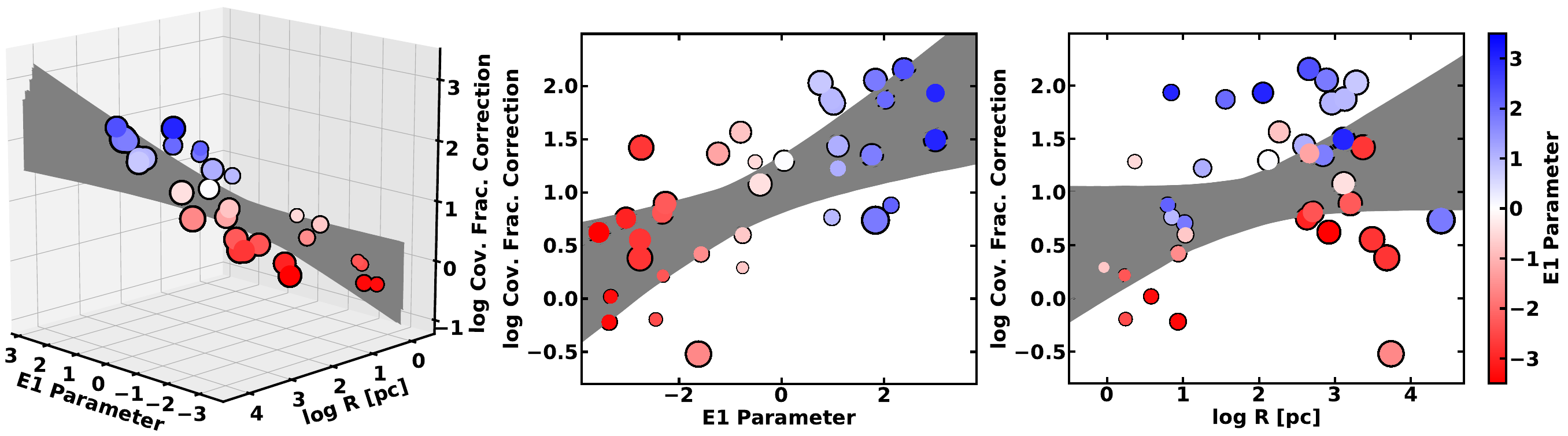}
\caption{The covering fraction corrections inferred from the
  regression analysis necessary to reconcile the observed [\ion{O}{3}]
  emission with that predicted to be emitted by the absorption line gas.
  The   assumed emission-line covering fraction was 0.1, so a log covering 
  fraction correction value of 1 would imply that the covering
  fraction of 0.01 would be needed to reconcile the values. The point
  color represents $E1$ and the point size represents $\log R$. {\it
    Left:} The multiparameter regression is a function of the E1
  parameter and $\log R$ and is therefore most accurately represented
  in 3D.   The bowtie
  surface shows the inferred errors from the {\tt mlinmix\_err}
  procedure; the results from all 48 combinations have been
  plotted. {\it Middle:} The results projected onto the $E1$ parameter
  plane.  {\it Right:} The results projected onto the $\log R$ plane.  
  \label{cov_frac_corr}}
\end{center}
\end{figure*}

We next examine how large the log covering fraction correction factors
are, and how they depend on the regression variables.  The results
from the regression are seen in Fig.~\ref{cov_frac_corr}.  The points
are plotted for all 48  combinations, and the grey shaded region shows
the 90\% confidence regions from the {\tt mlinmix\_err} procedure.
The left panel shows the results in three dimensions, while the
  right two panels show the two dimensional projections for each of
  the regression variables.
The assumed emission-line covering fraction was 0.1, so, for example,
a log covering fraction correction value of 1 would imply that the
covering fraction of 0.01 is needed to reconcile the observed and
predicted [\ion{O}{3}] values.  Many of the $E1>0$ objects 
have large covering fraction correction factors (1.5--2), which would
seem to imply that the emission-line gas covering fraction needs to be be very small
(0.001--0.003) in order to reconcile the observed and predicted
[\ion{O}{3}] emission.   

\begin{figure*}[!t]
\epsscale{1.0}
\begin{center}
\includegraphics[width=4.5truein]{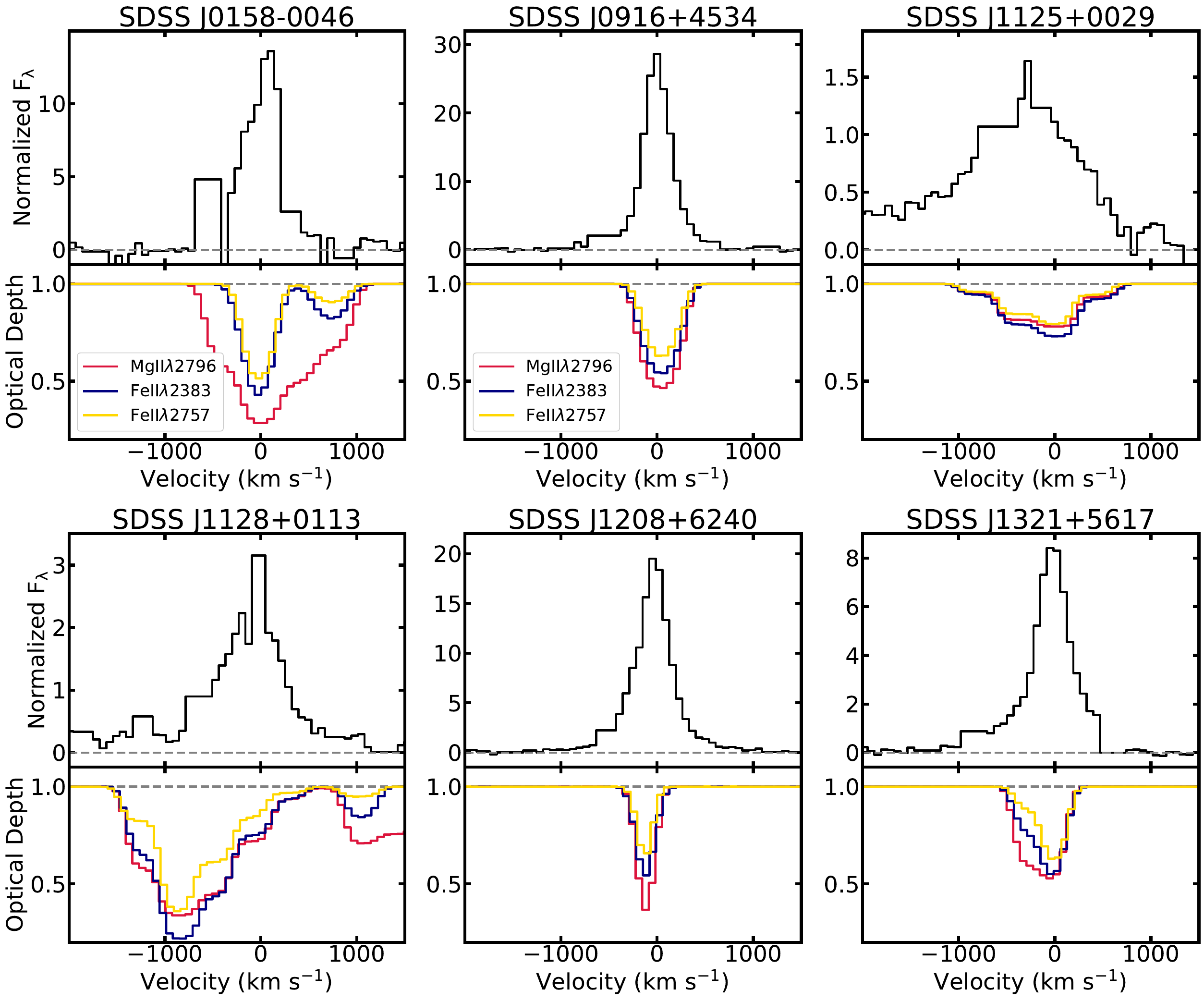}
\caption{Comparison of the [\ion{O}{3}]$\lambda 5008$ emission profile
  with the   absorption profile for objects with compact outflows
  ($\log R < 1$   [pc]) and  low log covering fraction correction 
  factor ($<0.5$).  {\it Top panel in each frame:} The
  [\ion{O}{3}]$\lambda 5008$ profile   from the normalized and
  continuum-subtraction spectrum.  {\it Bottom 
    panel in each frame:} The {\it SimBAL} derived opacity profiles
  for ground-state   \ion{Mg}{2}, ground-state \ion{Fe}{2}$\lambda
  2883$, and excited   state \ion{Fe}{2}$\lambda 2757$ taken from
  \citet[Fig.\ 13][]{choi22}.  
  \label{oiii_tau_vel}}
\end{center}
\end{figure*}

In contrast, several of the loitering outflow
objects \citep[$E1<0$ with a low-velocity and compact
  outflow][\S6.5, Fig.\ 18]{choi22} show very low covering fraction
  correction factors near zero, which means that the 
observed  [\ion{O}{3}] emission is consistent with being produced in
the outflow.  Fig.~\ref{oiii_tau_vel} compares the [\ion{O}{3}]
emission profile and the absorption opacity profiles \citep{choi22}
for the six objects with $\log R< 1$  and $\log$ 
covering fraction   correction less than 0.5. In these objects, the
data are roughly consistent   with the line emission and absorption
being produced in the same gas.     The profiles are not identical, but there
are consistent trends: objects with broader [\ion{O}{3}] emission lines
show broader absorption profiles.
For example, in SDSS~J1128+0113, the [\ion{O}{3}] line has a velocity width of
  $w_{80}=830\rm \, km\, s^{-1}$ \citep{leighly22} and the \ion{Mg}{2}
  absorption line has a velocity width of $3000 \rm \, km\, s^{-1}$
  \citep{choi22}.  In contrast, in SDSS~J0916$+$4534, the [\ion{O}{3}]
  line has a velocity width of
  $w_{80}=460\rm \, km\, s^{-1}$ \citep{leighly22} and the \ion{Mg}{2}
  absorption line has a velocity width of $500 \rm \, km\, s^{-1}$
  \citep{choi22}.  SDSS~J1321$+$5617 is particularly
interesting: both the [\ion{O}{3}] emission-line and the
absorption-line optical-depth profiles are narrow with a blue wing.

However, there is a flaw in this analysis.  We showed in \S~\ref{logr}
that the BAL outflow velocity in some objects was much smaller than
the local Keplerian velocity (Fig.~\ref{vkep_rat}).   The
[\ion{O}{3}] emission-line width in the same objects is also much less
than the Keplerian velocity, by factors of 5 to 24.  While we could
explain the small outflow velocity if the orientation is directly
along the line of sight, the same argument does not work for the
emission lines, since they are composed of emission from all lines of
sight.  

\begin{figure*}[!t]
\epsscale{1.0}
\begin{center}
\includegraphics[width=6.9truein]{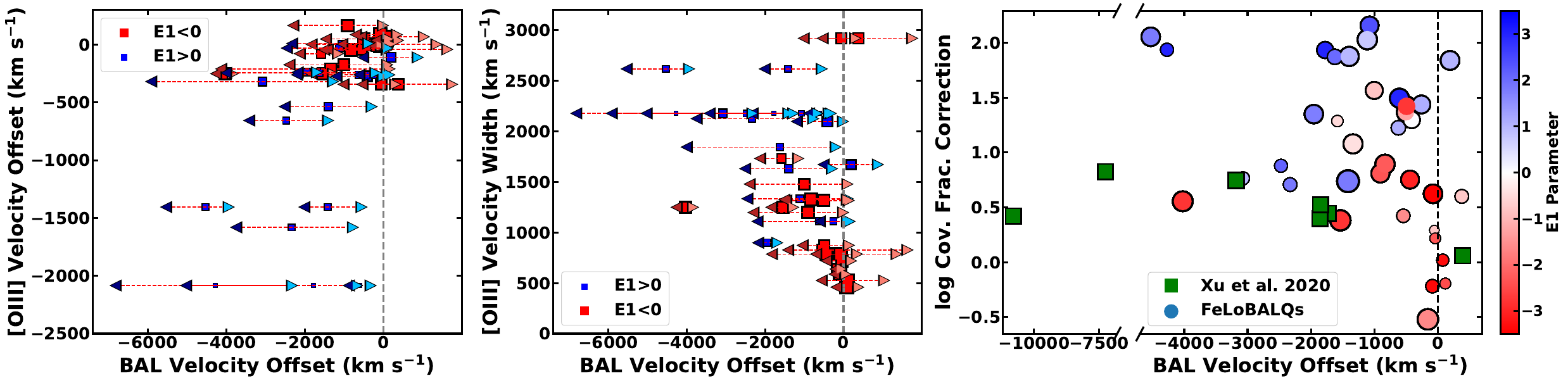}
\caption{Comparison of the BAL velocity offset with the [\ion{O}{3}]
  properties.   Note   that negative BAL velocities denote outflows.  In
  the  left and middle panels, the squares show
  the column-density weighted velocity, while the left and right
  triangles show the maximum and minimum speeds, respectively.  The
  size of the square scales with the log of the [\ion{O}{3}]
  equivalent width.   {\it 
    Left:}  There is a correlation between the [\ion{O}{3}] velocity
  offset and the BAL velocity offset ($p=1.9\times 10^{-4}$).
    {\it Middle:} There is an 
  anticorrelation between the [\ion{O}{3}] velocity
  width and the BAL velocity offset ($p=2.7\times 10^{-4}$).  {\it
    Right:} The covering   fraction correction factor is
  anti-correlated with the BAL velocity   offset for our sample
  ($p=7.8\times 10^{-3}$), but possibly also   for the \citet{xu20}
  sample. Larger  velocity offsets may produce   line emission that is
  distributed over a range of wavelength where   it may be blended
  with   the continuum or   hidden under strong \ion{Fe}{2} emission. 
  \label{bal_vel_vs}}
\end{center}
\end{figure*}

The correlation analysis discussed in \S~\ref{correlations}
indicates significant correlations between the BAL outflow velocity
offset and the [\ion{O}{3}] velocity offset and width
(Fig.~\ref{corr_optical_simbal}).  Fig.~\ref{bal_vel_vs} (left and
middle) explores these relationships.   In $E1>0$
objects, the [\ion{O}{3}] emission line is sometimes very small and
difficult to discern amid the sometimes strong and broad \ion{Fe}{2}
emission; therefore, points with the smallest symbols are less robustly 
measured.   The correlation between the BAL outflow velocity and
[\ion{O}{3}] velocity offset ($p=1.9\times 10^{-4}$) and the
anticorrelation between BAL outflow velocity and [\ion{O}{3}] velocity
width ($p=2.7\times 10^{-4}$) are apparent.

Our regression analysis above explored the relationships between the 
observed and predicted [\ion{O}{3}] luminosities for seven selected
measurements.  We also looked for relationships with other measured
parameters  via correlation analysis between the covering fraction
correction factor and the  {\it SimBAL} parameters
(Fig.~\ref{correlation_simbal}) and the optical parameters
(Fig.~\ref{corr_optical_simbal}).   Interestingly, although the
predicted [\ion{O}{3}] emission must increase with density, we found
no correlation between the covering fraction correction factor and the
BAL $\log n$.  This is explained by the fact that lower density BAL
gas is found at larger radii, where, for a fixed global covering
fraction of the line-emitting gas, the volume of emitting gas is
larger.  

One correlation stands out: the outflow velocity is anticorrelated
with the covering fraction correction factor.   This anticorrelation is
significant ($p<0.025$) in 27 of 48 cases using Spearman Rank; it did not
appear to be significant in the multiple regression analysis, as {\tt
  mlinmix\_err} uses  Pearson's R.  This dependence offers a way to
reconcile the large covering fraction correction factors required by
the $E1>0$ objects: the [\ion{O}{3}] emitted by the BAL could be so
broadened that it blends with the continuum, especially in low
signal-to-noise spectra or amidst strong \ion{Fe}{2} emission.

\section{Summary and Future Work}\label{discussion}

\subsection{Summary}\label{summary}

This is the third in a sequence of four papers that discuss the
properties of low-redshift FeLoBAL quasars.  Taken together, they
build a picture of the properties and physical conditions of the BAL
gas, and explore links between these properties to the accretion and
emission-line properties of the quasars.  

This paper combines the {\it SimBAL} \citep{choi22} and
optical emission-line analysis \citep{leighly22}.  The most
significant result is the discovery that the $E1$ parameter division
discovered by \citet{leighly22} 
carries over to the outflow properties (\S~\ref{logr}).   Among the
$E1>0$ high accretion rate objects, the outflow velocity {\it
  decreases} with distance from the central engine
(Fig.~\ref{logr_voffset}).  This is consistent with the expectation of
radiative line driving acceleration.  Among the $E1<0$ low accretion
rate objects, the outflow velocity {\it   increases} with the distance
from the central engine.  

We confirmed the relationship between the outflow velocity and
Eddington ratio previously reported by \citet{ganguly07}, i.e., at a
particular Eddington ratio, a range of velocities are observed up to a
maximum velocity which is itself a function of the Eddington ratio
(\S\ref{lbol}). Using the physical properties of the outflows obtained
from the {\it   SimBAL} analysis, we investigated whether the scatter
in the velocity at a particular Eddington ratio could be a consequence
of a scatter in the force multiplier or the launch radius.  We found that
among $E1>0$ objects, the scatter in the velocity could plausibly be
attributed to a scatter in the launch radius, while among $E1<0$
objects, the scatter in the velocity could be attributed to scatter in
the force multiplier.  
 
We investigated the volume filling factor of the outflows, both for
the full sample (\S~\ref{full}) and for the $z<1$ subsample
(\S~\ref{covfrac}).  The full sample reveals a large range of $\log$
volume filling factors, from $-6$ to $-1$.   At large distances from
the central engine, the $\log$ volume filling factors were less than
$-3$, similar to those inferred in HiBAL quasars
\citep[e.g.,][]{gabel06,hamann11,hamann13}.  Closer to the black hole,
for $\log R \lesssim 1$,  the full range of volume filling factors
was found.  We also found that the special BAL classes identified by
\citet{choi22} (the loitering outflows and the other overlapping
trough objects) are also divided by their $E1$ parameter, and by
extension, their Eddington ratio.  The loitering outflow objects have
$E1<0$ and  low Eddington ratios, while the other overlapping trough
objects have $E1>0$ and high Eddington ratios.  Moreover, although
both special types of   
FeLoBAL quasar outflows are compact, with typical location $\sim 10$
parsecs from the central engine, a dramatically different number of
assumed spherical clouds would be required to occult the continuum
emission region  (Fig.~\ref{numclouds_logr}).  For the loitering
outflows, a single cloud (or continuous outflow) would be sufficient.
For the other overlapping trough absorbers, 100s to 1000s of clouds
would be required. Of course, the outflow may not have the structure
of discrete spherical clouds; the point is that the differing physical
conditions of these two categories of outflows tell us that the
structure of the outflows is dramatically different.     

We also investigated the relationship between the observed
[\ion{O}{3}] emission line and the [\ion{O}{3}] emission predicted to
be produced by the BAL outflow gas (\S\ref{oiiiemission}).   This
analysis first underlines the very large range of equivalent widths
observed in this sample.  The observed [\ion{O}{3}] luminosity and the
estimated bolometric 
luminosity both span about 1.6 dex, but the ratio of the two
luminosities spans 2.1 dex. We found that in order to 
reconcile the observed and predicted $\log$ [\ion{O}{3}] luminosities,
the 
emission-line gas global covering fraction may depend on the $E1$
parameter and the location of the outflow $\log R$ (Fig.~\ref{lum_vs_lum}).
At the same time, the covering fraction correction factor (defined as
the difference between the observed and predicted [\ion{O}{3}]
luminosity) was observed to be correlated with the outflow velocity,
which may imply that the predicted strong [\ion{O}{3}] emission in
$E1>0$ objects is broadened and hidden under their typically strong
\ion{Fe}{2} emission (Fig.~\ref{bal_vel_vs}).  Most intriguing were
the six $E1<0$ objects with the lowest covering fraction correction
factors, whose [\ion{O}{3}] profiles resembled the BAL absorption
profiles (Fig.~\ref{oiii_tau_vel}).  

The final paper in this series of four papers (Leighly et al.\ in
prep.) includes an analysis of the the broad-band optical/IR
properties and discusses the potential implications for quasar
evolution scenarios.    

\subsection{Future Work}\label{future}

This sample was limited to objects with redshifts less than 1.63 for
the full 50-object sample from \citet{choi22} and
less than 1.0 for the optical emission-line analysis from
\citet{leighly22}.  FeLoBAL quasars  can be observed up to $z\sim 3$
in ground-based optical-band spectra.  Because the SDSS is a 
flux-limited survey, we generally 
expect higher-redshift objects to be more luminous; a sample currently
being analyzed has bolometric luminosities about one order of magnitude
larger than the $z<1$ sample.  Assuming a similar distribution of
Eddington ratios, the higher-redshift objects will have larger black
hole masses, and therefore a softer (more UV-dominant) SED.  The
softer SED could influence the properties of the BAL outflows in two
ways: 1.\ the ions present in the outflowing gas would be different,
e.g., tend toward lower ionization species \citep[e.g.,][]{leighly07}, 
and 2.\ the velocities might be larger, since the SED will produce a
relatively larger number of UV photons that can transfer momentum by
resonance scattering in the outflowing gas.  Preliminary analysis of a
higher-redshift sample that is  being observed in the near infrared   
shows evidence of higher outflow velocities \citep{voelker21}.
Another interesting feature of the high-redshift 
sample is that the larger redshifts provide access to the
high-ionization lines (e.g., \ion{C}{4}, \ion{Si}{4}) that these
objects share with the much more common HiBAL quasars.  Preliminary
analysis shows that HiBAL quasars seem to be much different than the 
FeLoBAL quasars \citep{hazlett19, leighly_aas19}; in particular, the
high ionization lines sometimes have complicated velocity structure
and certainly extend to higher velocities.  The 
potential link between the low-ionization line and high-ionization
line properties may also prove to be very illuminating.  

Our investigation of whether there could be a relationship between the
ionized outflows manifest in [\ion{O}{3}] emission and the BAL
outflows showed that in most objects such a relationship would be
possible only if the emission-line covering fraction is extremely low,
or if the [\ion{O}{3}] is broad and blended with the \ion{Fe}{2}
emission.  An exception was several of the loitering outflow objects,
where the emission line and absorption line profiles appeared to
resemble one another.  All of these objects had compact outflows, so
the possibility that there is a direct relationship might be able to be
tested with spatially-resolved  observations of the [\ion{O}{3}]
emission; unlike many quasars
\citep[e.g.,][]{harrison14,bischetti17,vayner21b}, the [\ion{O}{3}] 
emission in these objects should be unresolved if it does indeed
originate in the BAL outflow.     Proving a relationship would be very
difficult in general.  However, we might be able to falsify one.    At
a redshift of 0.9, the angular scale is about 7.9 kpc per arc second,
so it would be possible to determine whether the [\ion{O}{3}] emission
was extended or compact on the scale of a kilo-parsec.   Many of the 
{\it SimBAL} solutions indicate BAL outflows with size scales much
less than one kiloparsec.  The [\ion{O}{3}] emission should be
unresolved in such objects.
\vskip 1pc

Regardless of the details, the analysis presented here has shown that
again, key physical properties of the outflows differ as a
function of the location in the quasar, and as a function of accretion
rate as probed by the $E1$ parameter.  These patterns and differences
have broad potential implications.  We may finally be able to
understand the acceleration mechanisms that operate in quasars. In
addition, we may be able to use the accretion properties measured from
the emission lines to statistically infer the outflow properties in
objects that do not show outflows along the line of sight.  These
ambitious goals will require much more work and analysis of many more
objects, but at least we have identified a promising path  forward.   

\acknowledgements

We thank the current and past {\it SimBAL} group members and the anonymous referee for useful discussions and comments on the manuscript.
Support for {\it SimBAL} development and analysis is provided by NSF
Astronomy and Astrophysics Grants No.\ 1518382 and 2006771. This work
was performed in part at Aspen Center for Physics, which is supported
by National Science Foundation grant PHY-1607611. SCG thanks the
Natural Science and Engineering Research Council of Canada.

Long before the University of Oklahoma was established, the land on
which the University now resides was the traditional home of the
“Hasinais” Caddo Nation and “Kirikiris” Wichita \& Affiliated
Tribes. This land was also once part of the Muscogee Creek and
Seminole nations.

We acknowledge this territory once also served as a hunting ground,
trade exchange point, and migration route for the Apache, Comanche,
Kiowa and Osage nations. Today, 39 federally-recognized Tribal nations
dwell in what is now the State of Oklahoma as a result of settler
colonial policies designed to assimilate Indigenous peoples.

The University of Oklahoma recognizes the historical connection our
university has with its Indigenous community. We acknowledge, honor
and respect the diverse Indigenous peoples connected to this land. We
fully recognize, support and advocate for the sovereign rights of all
of Oklahoma’s 39 tribal nations.

This acknowledgement is aligned with our university’s core value of
creating a diverse and inclusive community. It is our institutional
responsibility to recognize and acknowledge the people, culture and
history that make up our entire OU Community.

Funding for the SDSS and SDSS-II has been provided by the Alfred
P. Sloan Foundation, the Participating Institutions, the National
Science Foundation, the U.S. Department of Energy, the National
Aeronautics and Space Administration, the Japanese Monbukagakusho, the
Max Planck Society, and the Higher Education Funding Council for
England. The SDSS Web Site is http://www.sdss.org/.

The SDSS is managed by the Astrophysical Research Consortium for the
Participating Institutions. The Participating Institutions are the
American Museum of Natural History, Astrophysical Institute Potsdam,
University of Basel, University of Cambridge, Case Western Reserve
University, University of Chicago, Drexel University, Fermilab, the
Institute for Advanced Study, the Japan Participation Group, Johns
Hopkins University, the Joint Institute for Nuclear Astrophysics, the
Kavli Institute for Particle Astrophysics and Cosmology, the Korean
Scientist Group, the Chinese Academy of Sciences (LAMOST), Los Alamos
National Laboratory, the Max-Planck-Institute for Astronomy (MPIA),
the Max-Planck-Institute for Astrophysics (MPA), New Mexico State
University, Ohio State University, University of Pittsburgh,
University of Portsmouth, Princeton University, the United States
Naval Observatory, and the University of Washington.

Funding for SDSS-III has been provided by the Alfred P. Sloan
Foundation, the Participating Institutions, the National Science
Foundation, and the U.S. Department of Energy Office of Science. The
SDSS-III web site is http://www.sdss3.org/.

SDSS-III is managed by the Astrophysical Research Consortium for the
Participating Institutions of the SDSS-III Collaboration including the
University of Arizona, the Brazilian Participation Group, Brookhaven
National Laboratory, Carnegie Mellon University, University of
Florida, the French Participation Group, the German Participation
Group, Harvard University, the Instituto de Astrofisica de Canarias,
the Michigan State/Notre Dame/JINA Participation Group, Johns Hopkins
University, Lawrence Berkeley National Laboratory, Max Planck
Institute for Astrophysics, Max Planck Institute for Extraterrestrial
Physics, New Mexico State University, New York University, Ohio State
University, Pennsylvania State University, University of Portsmouth,
Princeton University, the Spanish Participation Group, University of
Tokyo, University of Utah, Vanderbilt University, University of
Virginia, University of Washington, and Yale University.

\software{Cloudy \citep{ferland13}, mlinmix\_err
  \citep{kelly07},  SimBAL \citep{leighly18}}



\end{CJK}
\end{document}